\documentclass{ws-ijmpa}
\usepackage[super,compress]{cite}
\usepackage{graphicx}
\usepackage{braket}
\usepackage{bm}
\usepackage{comment}
\usepackage{here}
\usepackage{multirow}
\usepackage{color}

\def\Journal#1#2#3#4{{#1} {\bf #2}, #3 (#4)}
\def\Journal#1#2#3#4{{#1} {\bf #2}, #3 (#4)}

\def\AandA{Astron. Astrophys.}

\def\APP{Astropart. phys.}

\def\IJMPA{Int. J. Mod. Phys. A}
\def\IJMPD{Int. J. Mod. Phys. D}

\def\JCAP{J. Cosmol. Astropart. Phys.}
\def\JHEP{J. High Energy Phys.}

\def\MNRAS{Mon. Not. R. Astron. Soc.}
\def\MPLA{Mod. Phys. Lett. A}

\def\PDU{Phys. Dark. Univ.}

\def\PLB{{Phys. Lett.} B}

\def\PRL{Phys. Rev. Lett.}
\def\PRD{Phys. Rev. D}

\def\RPP{Rep. Prog. Phys.}

\def\UNIV{Universe}

\begin{document}
\markboth{T. Kitabayashi and A. Shimizu}{Running scalar spectral index in warm natural inflation}

%
\catchline{}{}{}{}{}
%


\title{Running scalar spectral index in warm natural inflation}

\author{Teruyuki Kitabayashi\footnote{Corresponding author}}

\address{Department of Physics, Tokai University,\\
4-1-1 Kitakaname, Hiratsuka, Kanagawa 259-1292, Japan\\
teruyuki@tokai.ac.jp}

\author{Aya Shimizu}

\address{Graduate School of Science, Tokai University,\\
4-1-1 Kitakaname, Hiratsuka, Kanagawa 259-1292, Japan}

\maketitle

\begin{history}
\received{Day Month Year}
\revised{Day Month Year}
\end{history}

\begin{abstract}
The validity of inflation models is mainly evaluated according to the consistency of the predicted scalar spectral index $n_{\mathrm{s}}$, the tensor scalar ratio $r$, and the running scalar spectral index $\alpha_{\mathrm{s}}$ with cosmic microwave background observations. In warm inflation (WI) scenarios, one can find exact analytical solutions for $\alpha_{\mathrm{s}}$ in principle, but long expressions may be obtained. Previous studies for WI scenarios have only shown approximate analytical solutions or numerical results for $\alpha_{\mathrm{s}}$. In this study, we present a general analytical expression of $\alpha_{\mathrm{s}}$ without approximation in WI. By providing an analytical expression, even if it is mathematically redundant, we believe that $\alpha_{\mathrm{s}}$ will be studied across a broader range of WI models in the future. The obtained analytical expression of $\alpha_{\mathrm{s}}$ is used in the study of warm natural inflation (WNI). Although $n_{\mathrm{s}}$ and $r$ have been previously investigated, $\alpha_{\mathrm{s}}$ is omitted in previous studies on WNI. Our study of $\alpha_{\mathrm{s}}$ completes previous phenomenological studies on WNI. In particular, the lower limit of the symmetry-breaking scale in WNI becomes more concrete in this study.
\end{abstract}



\section{Introduction\label{section:introduction}}
The inflationary paradigm \cite{Guth1981PRD,Sato1981MNRAS,Linde1982PLB,Albrecht1982PRL} is a cornerstone of standard cosmology, and several inflationary models consistent with observations have been developed \cite{Martin2014PDU}. In traditional cold inflation (CI) scenarios, the inflaton is assumed to have no interaction with other fields during inflation. In contrast, in warm inflation (WI) scenarios \cite{Berera1995PRD74,Berera1995PRD75,Dymnikova2001MPLA,Berera2009RPP,GIL2009IJMPA}, the inflaton interacts with other fields during inflation, and radiation production occurs. Because general relativity only requires that the inflaton energy density dominates for inflation to occur, substantial radiation energy densities present during inflation are allowed in principle. The validity of both CI and WI is mainly evaluated by the consistency of the predicted scalar spectral index $n_{\mathrm{s}}$ and the tensor scalar ratio $r$ with cosmic microwave background (CMB) observations.

In addition, the running scalar spectral index $\alpha_{\mathrm{s}}$ can be used to further check inflation models' consistency. In CI scenarios, $\alpha_{\mathrm{s}}$ can be expressed in simple mathematical expressions using slow-roll parameters. In contrast, in WI scenarios, one can find exact analytical solutions for $\alpha_{\mathrm{s}}$ in principle but may get long expressions \cite{Das2023Universe}. Therefore, previous studies on WI have only shown approximate analytical solutions \cite{Das2023Universe,Trivedi2024APP} or numerical results \cite{Benetti2017PRD,Montefalcone2024JCAP,Rodrigues2025arXiv} for $\alpha_{\mathrm{s}}$. Moreover, whereas $\alpha_{\mathrm{s}}$ has been studied in several CI models, it has been studied in only limited WI models, such as warm chaotic inflation, warm hilltop inflation, warm Higgs-like inflation, and minimal WI \cite{Benetti2017PRD,Das2023Universe,Montefalcone2024JCAP}.

The objectives of this study are outlined below:
\begin{itemize}
\item The first objective is showing a general analytical expression of $\alpha_{\mathrm{s}}$ without approximation for two major dissipative coefficients: linear and cubic. We also demonstrate that the analytical expressions of $\alpha_{\mathrm{s}}$ is consistent with the results from numerical calculation in Ref. \cite{Rodrigues2025arXiv}. By providing an analytical expression, even if it is mathematically redundant, we believe that $\alpha_{\mathrm{s}}$ will be studied across a broader range of WI models in the future.

\item The second goal of this study is to evaluate the running scalar index $\alpha_{\mathrm{s}}$ in the warm natural inflation (WNI) model \cite{Mohanty2008PRD,Visinelli2011JCAP,Mishra2012PLB,Berghaus2020JCAP,AlHallak2023Universe,Montefalcone2023JCAP,Reyimuaji2021JCAP,Mukuno2024PRD,Correa2022PLB,Correa2024PRD}. Although the cold natural inflation (CNI) model \cite{Freese1990PRL,Adams1993PRD,Freese2008IJMPD,Freese2015JCAP} is appealing because of its well-motivated origin of the inflaton potential, it requires a large symmetry breaking scale \cite{Martin2014PDU}. This is theoretically undesirable because it may cause large gravitational corrections to the potential. Therefore, modified CNI models \cite{Ross2010PLB,Ross2010PLB2,Hebecker2013PRD,Gonzalez2014PLB,Carone2014PRD,Vazquez2015JCAP,Ross2016JHEP,German2017JCAP,Ali2017PRD,Gong2021PRD,Kitabayashi2023PRD} and WNI models have been proposed. Although $n_{\mathrm{s}}$ and $r$ in WNI have already been studied \cite{Mohanty2008PRD,Visinelli2011JCAP,Mishra2012PLB,Berghaus2020JCAP,AlHallak2023Universe,Montefalcone2023JCAP,Reyimuaji2021JCAP,Mukuno2024PRD,Correa2022PLB,Correa2024PRD}, there have been no studies on $\alpha_{\mathrm{s}}$ in the WNI scenario. This work aims to fill this gap and complete the phenomenological study of WNI.
\end{itemize}

The remainder of this article is organized as follows. In Section \ref{section:WI}, we provide an overview of the WI scenario. Then, we present a general and exact analytical solution of $\alpha_{\mathrm{s}}$ for two major dissipative coefficients in the WI scenario (this is the first purpose in this study) in Section \ref{section:exact_solutions}. In Section \ref{section:WNI}, we evaluate $\alpha_{\mathrm{s}}$, $n_{\mathrm{s}}$, and $r$ in the WNI model (this is the second purpose in this study). The article concludes with a summary of the key findings in Section \ref{section:summary}.

\section{Warm inflation \label{section:WI}}
\subsection{Field evolution}
In the WI scenario, the equation of motion of the inflaton $\phi$ in the Friedmann--Robertson--Walker metric is given as follows:
\begin{equation}
\ddot{\phi} +(3H + \Gamma)\phi + V_{,\phi} = 0,
\end{equation}
where $H$ and $\Gamma$ denote the Hubble parameter and the dissipative coefficient, respectively. The dot indicates the derivation with respect to cosmic time, and we use the notation $V_{,\phi}=\partial V/ \partial \phi$.

As usual, we assume that the thermalization timescale of the radiation is much shorter than $1/\Gamma$ so that the energy density of the radiation is given by
\begin{equation}
\rho_r = \frac{\pi^2}{30}g_*T^4 = \tilde{g}_*T^4,
\label{Eq:rho_r}
\end{equation}
where $g_*$ is the effective number of degrees of relativistic particles at temperature $T$. From the energy conservation, the evolution of the radiation energy density can be obtained as follows:
\begin{equation}
\dot{\rho}_r+ 4H\rho_r = \Gamma \dot{\phi}^2.
\end{equation}
The term $\Gamma \dot{\phi}^2$ represents the energy transfer from the inflaton to the radiation bath. The Friedmann equation for the background evolution is given as follows:
\begin{equation}
H^2 = \frac{1}{3M_{\rm pl}^2} \left( \frac{1}{2}\dot{\phi}^2 + V + \rho_r \right),
\end{equation}
where $M_{\rm pl} = 1/\sqrt{8\pi G}$ is the reduced Planck mass ($G$ denotes the gravitational constant).

Inflation is realized when the Hubble expansion rate $H$ is approximately constant. This phenomenon occurs when the potential $V$ is approximately flat and the potential energy dominates all other forms of energy. In this case, the usual slow-roll approximation of the inflation potential can be used. In the slow-roll regime of the inflaton field, higher-order derivatives can be neglected:
\begin{equation}
\ddot{\phi} \ll H\dot{\phi}, \quad \dot{\rho}_r \ll H\rho_r.
\end{equation}
Consequently, the equation of motion for the inflaton is as follows:
\begin{equation}
\dot{\phi} \simeq -\frac{V_{,\phi}}{3H+\Gamma} = -\frac{V_{,\phi}}{3H(1+Q)},
\label{Eq:phi_dot}
\end{equation}
where the dimensionless dissipative ratio
\begin{equation}
Q=\frac{\Gamma}{3H}.
\end{equation}
The equation of motion of the radiation bath and the Friedmann equation for the background evolution are as follows:
\begin{equation}
4H\rho_r \simeq \Gamma \dot{\phi}^2, \quad  \left(\rho_r \simeq \frac{3Q\dot{\phi}^2}{4} \right)
\label{Eq:EoM}
\end{equation}
and 
\begin{equation}
H^2 \simeq \frac{V}{3M_{\rm pl}^2},
\label{Eq:H2}
\end{equation}
respectively.

The consistency of the inflation models can be estimated using the following slow-roll parameters:
\begin{equation}
\epsilon = \frac{M_{\rm pl}^2}{2}\left( \frac{V_{,\phi}}{V}\right)^2, \quad \eta = M_{\rm pl}^2  \frac{V_{,\phi\phi}}{V}, \quad \beta = M_{\rm pl}^2 \frac{\Gamma_{,\phi} V_{,\phi}}{\Gamma V}.
\end{equation}
The slow roll conditions are given by the following:
\begin{equation}
\epsilon < 1 + Q, \quad \eta < 1+Q, \quad \beta <1+Q.
\end{equation}
The following equations showing alternative definitions of the slow-roll parameters are useful for determining the era of the WI end:
\begin{equation}
\epsilon_w =  \frac{\epsilon}{1+Q} , \quad
\eta_w = \frac{\eta}{1+Q} , \quad
\beta_w = \frac{\beta}{1+Q}.
\end{equation}
WI ends when
\begin{equation}
\epsilon_w = 1, \quad {\rm or} \quad  |\eta_w| = 1 \quad  {\rm or} \quad  |\beta_w| = 1.
\label{Eq:condition_of_end_of_inflation}
\end{equation}
%

\subsection{Power spectrum}
The primordial scalar power spectrum $\mathcal{P}_R(k)$ for curvature perturbations can be expanded in terms of the scalar spectral index (spectral tilt) $n_{\mathrm{s}}$ and the running of the scalar spectral index $\alpha_{\mathrm{s}}$ as
\begin{equation}
\mathcal{P}_R(k) = \mathcal{P}_*\left(\frac{k}{k_*}\right)^{n_{\mathrm{s}}-1+\frac{\alpha_{\mathrm{s}}}{2}\ln(k/k_*)},
\end{equation}
where $k$ denotes the wave number, $\mathcal{P}_*$ is the amplitude of the scalar power spectrum, and $k_*$ is a pivot scale relevant for the CBM observation scale; that is, $k_*=0.05$ Mpc$^{-1}$ for $\mathcal{P}_* \simeq 2.1 \times 10^{-9}$ \cite{PLANCK2020AA_X}. The suffix $*$ indicates the quantities evaluated at the Hubble radius crossing of the pivot scale $k_*$.

In the CI, the quantum fluctuation of the inflaton field is the only source of the primordial power spectrum. In the WI, the presence of dissipation affects the background dynamics of the inflaton field as well as its perturbations. Because of the temperature dependence of the dissipation coefficient, the inflaton and radiation fluctuations are coupled. The inflaton can be excited and obeys a Bose--Einstein distribution. When $T>H$, thermal fluctuations can dominate quantum density fluctuations. Consequently, the additional thermal effects significantly impact the primordial power spectrum. The dimensionless primordial scalar curvature power spectrum with the thermal effects is obtained as \cite{Hall2004PRD, Gil2011JCAP,Graham2009JCAP,Ramos2013JCAP,Ballesteros2022JCAP,Ballesteros2024JCAP}
\begin{equation}
\mathcal{P}_R =  \left(\frac{H^2}{2\pi \dot{\phi}}\right)^2\left(1+2n_{\rm BE} + \frac{2 \sqrt{3}\pi Q}{\sqrt{3+4\pi Q}}\frac{T}{H} \right)  G(Q),
\end{equation}
where all quantities are evaluated at horizon crossing. $n_{\rm BE} = 1/ [\exp(H/T)-1]$ denotes the inflaton statistical distribution due to the presence of the radiation bath. As commonly assumed, the inflatons are used to follow a Bose--Einstein radiation distribution. The function $G(Q)$ accounts for the coupling of the inflaton and radiation fluctuations due to a temperature-dependent dissipation rate. As shown later, $G(0) = 1$. In the limit $Q \rightarrow 0$ and $T\rightarrow 0$, the scalar curvature power spectrum in the CI scenario can be obtained.

The scalar spectral index $n_{\mathrm{s}}$ can be obtained at the horizon crossing ($k=k_*)$ as follows:
\begin{equation}
n_{\mathrm{s}}-1 = \lim_{k \rightarrow k_*} \frac{d \ln \mathcal{P}_R}{d \ln (k/k_*)}  \equiv \frac{d \ln \mathcal{P}_R}{d \ln k}\simeq \frac{d \ln \mathcal{P}_R}{d N},
\end{equation}
where $dN = Hdt$ denotes the differential increment in the number of e-folds. The general expression of the scalar spectral index is given by \cite{Reyimuaji2021JCAP,Montefalcone2023JCAP}
\begin{align}
n_{\mathrm{s}}-1 &=  4\frac{d \ln H}{dN} - 2\frac{d \ln \dot{\phi}}{dN}  + \left(1+2n_{\rm BE} + \frac{2 \sqrt{3}\pi Q}{\sqrt{3+4\pi Q}}\frac{T}{H} \right)^{-1} \nonumber \\
&\times \left\{ 2n_{\rm BE}^2 e^{H/T}\frac{H}{T}\left(\frac{d \ln T}{d N} - \frac{d\ln H}{dN} \right) \right. \nonumber \\
& + \left. \frac{2 \sqrt{3}\pi Q}{\sqrt{3+4\pi Q}}  \frac{T}{H} \left[ \frac{3+2\pi Q}{3+4\pi Q}  \frac{d\ln Q}{dN} + \frac{d\ln T}{d N} - \frac{d\ln H}{d N} \right] \right\} + \frac{Q}{G(Q)}\frac{dG(Q)}{dQ}\frac{d\ln Q}{dN},
\label{Eq:ns_1}
\end{align}
where
\begin{align}
\frac{T}{H}=\left( \frac{9}{4\tilde{g}_*} \frac{Q}{(1+Q)^2} \frac{M_{\rm pl}^6 V_{,\phi}^2}{V^3} \right)^{1/4}.
\end{align}

The running scalar spectral index can be expressed as follows:
\begin{equation}
\alpha_{\mathrm{s}} = \lim_{k \rightarrow k_*} \frac{d n_{\mathrm{s}}}{d \ln (k/k_*)}  \equiv \frac{d n_{\mathrm{s}}}{d \ln k} \simeq \frac{d n_{\mathrm{s}}}{d N}.
\end{equation}
The general expression of the running scalar spectral index is lengthy. The general expression for the two commonly discussed dissipation coefficients is shown later.

Owing to the weakness of gravitational interaction, we can assume that the tensor power spectrum is not affected by the dissipative dynamics \cite{Moss2008JCAP, Bhattacharya2006PRL, Ramos2013JCAP,Montefalcone2023JCAP}. Therefore, the tensor power spectrum in the WI is unchanged with respect to the spectrum in the CI scenario:
\begin{equation}
\mathcal{P}_T = \frac{2H^2}{\pi^2M_{\rm pl}^2}.
\end{equation}

The tensor-to-scalar ratio $r$ is the same as in the CI case. The general formula for the tensor-to-scalar ratio is obtained as \cite{Montefalcone2023JCAP}
\begin{equation}
r = \frac{\mathcal{P}_T}{\mathcal{P}_R} = \frac{16 \epsilon}{(1+Q)^2}\left(1+2n_{\rm BE} + \frac{2 \sqrt{3}\pi Q}{\sqrt{3+4\pi Q}}\frac{T}{H} \right)^{-1} \frac{1}{G(Q)}.
\end{equation}
In the limit $Q \rightarrow 0$ and $T \rightarrow 0$, the relation in the CI $r=16\epsilon$ is recovered. Because of the enhancement of the term
\begin{equation}
\left(1+2n_{\rm BE} + \frac{2 \sqrt{3}\pi Q}{\sqrt{3+4\pi Q}}\frac{T}{H} \right)G(Q),
\end{equation}
in the scalar curvature power spectrum $\mathcal{P}_R$, WI can decrease the tensor--scalar ratio through dissipative and thermal effects.

\subsection{Dissipation}
We employ the commonly used dissipation rate parametrization
\begin{equation}
\Gamma = C\frac{T^p}{f_1^{p-1}},
\label{Eq:GammaTf}
\end{equation}
where $C$ is a dimensionless factor and $f_1$ denotes a constant \cite{Moore2011JHEP,Berghaus2020JCAP,Mishra2012PLB,Reyimuaji2021JCAP,Montefalcone2023JCAP}. A different parametrization $\Gamma = CT^p/\phi^{p-1}$ has also been used \cite{Ramos2013JCAP,Berghaus2020JCAP,Montefalcone2023JCAP}. The exact expressions of $n_{\mathrm{s}}$ and $\alpha_{\mathrm{s}}$, which are shown later, are available not only for $\Gamma = CT^p/f_1^{p-1}$ but also for $\Gamma = CT^p/\phi^{p-1}$.

To determine the expression for $n_{\mathrm{s}}$ and $\alpha_{\mathrm{s}}$, the following derivatives for $\Gamma = \Gamma(\phi, T)=CT^p/\phi^{p-1}$ in terms of the slow-roll parameters are useful \cite{Ramos2013JCAP,Berghaus2020JCAP,Montefalcone2023JCAP}.
\begin{align}
\frac{d \ln H}{dN} &= -\frac{\epsilon}{1+Q}, \\
\frac{d\ln\phi}{dN} & = -\frac{\sigma}{1+Q}, \\
\frac{d \ln Q}{dN} &=\frac{2\left[(p+2)\epsilon - p\eta  + 2 (p-1)\sigma \right]}{4-p+Q(p+4)}, \\
\frac{d \ln T}{dN} &= \frac{(3+Q)\epsilon - 2(1+Q)\eta + (1-Q)(p-1)\sigma}{(1+Q)[4-p+Q(p+4)]},   \\
\frac{d \ln \dot{\phi}}{dN} &= \frac{ [4-p(1+Q)]\epsilon + (p-4)(1+Q)\eta - 4Q(p-1)\sigma }{(1+Q)[4-p+Q(p+4)]},
\end{align}
where
\begin{equation}
\sigma = \frac{\Gamma}{\phi \Gamma_{,\phi}}\beta = M_{\rm pl}^2 \frac{V_{,\phi}}{\phi V}.
\end{equation}
For $\Gamma = CT^p/f^{p-1}$, we obtain $\sigma = \beta = 0$ as $\Gamma_{,\phi} = 0$.

For the dissipation coefficient $\Gamma = CT^p/f_1^{p-1}$, the general relation between the dissipation rate $Q$ and the inflaton potential $V$ is obtained as \cite{Reyimuaji2021JCAP,Montefalcone2023JCAP}
\begin{equation}
Q^{4-p}(1+Q)^{2p} = \frac{C^4 M_{\rm pl}^{2(p+2)}}{9 \cdot 4^p \tilde{g}_*^p} \frac{1}{f_1^{4(p-1)}} \frac{(V_{,\phi})^{2p}}{V^{p+2}}.
\label{Eq:Q_V}
\end{equation}

For later use, we define
\begin{align}
A =\left(1+2n_{\rm BE} + \frac{2 \sqrt{3}\pi Q}{\sqrt{3+4\pi Q}}\frac{T}{H} \right)^{-1}, \quad
B =n_{\rm BE}^2 e^{H/T}\frac{H}{T}+ \frac{6\sqrt{3}\pi Q (1+\pi Q)}{(3+4\pi Q)^{3/2}} \frac{T}{H},
\end{align}
and
\begin{align}
D&=4n_{\rm BE}^2 e^{H/T}\frac{H}{T} \left[ 2(1+2Q)\epsilon - (1+Q)\eta +(1-Q)\sigma \right] \nonumber \\
&+ \frac{4\sqrt{3} \pi Q}{(3+4\pi Q)^{3/2}} \frac{T}{H} \left[ (21 + 9 (3 + 2 \pi) Q + 26 \pi Q^2)\epsilon \right.  \nonumber \\
& \quad \left. -2(1 + Q) (6 + 5\pi Q)  \eta  + (15 + 3 (3 + 4 \pi) Q + 4\pi Q^2) \sigma \right]. 
\end{align}
In addition, the following derivatives are used later:
\begin{align}
\frac{d\ln A}{dN} =&  -A\left\{2\cdot\frac{d n_{\rm BE}}{dN} + \frac{2\sqrt{3} \pi Q}{\sqrt{3+4\pi Q}} \frac{T}{H} \left(\frac{3+2\pi Q}{3+4\pi Q} \frac{d\ln Q}{dN}+ \frac{d\ln T}{dN}-\frac{d\ln H}{dN}\right)\right\},   \nonumber \\
\frac{dn_{\rm BE}}{dN} =& n_{\rm BE}^2 e^{H/T} \frac{H}{T}\left( \frac{d\ln T}{dN}-\frac{d\ln H}{dN} \right), \quad
\frac{d}{dN}\frac{dG}{dQ} = Q \frac{d\ln Q}{dN}\frac{d^2G}{dQ^2}, \nonumber \\
\quad \frac{d\ln G}{dN} =& \frac{Q}{G} \frac{dG}{dQ}\frac{d\ln Q}{dN}, 
\end{align}
and
\begin{align}
\frac{d\epsilon}{dN} = \frac{2\epsilon}{1+Q}(2\epsilon - \eta), \quad
\frac{d\eta}{dN} = \frac{\epsilon}{1+Q}(2\eta - \xi), \quad
\frac{d\sigma}{dN} = \frac{\sigma}{1+Q}(2\epsilon - \eta + \sigma),
\end{align}
where
\begin{equation}
\xi = 2M_{\rm pl}^2 \frac{V_{,\phi\phi\phi}}{V_{,\phi}}.
\end{equation}
%

\subsection{Number of e-folds \label{section:N}}
The expansion rate of the universe in the inflation epoch is parameterized by the number of e-folds, defined as follows:
\begin{align}
N = \ln \left( \frac{a_{\rm end}}{a_k} \right) = \int_{t_k}^{t_{\rm end}} Hdt,
\end{align}
where $a_{\rm end}$ and $a_k$ are the scale factors when the inflation ends and the scale $k$ crosses the horizon $(k=a_kH)$, respectively. In WI, we obtain
\begin{align}
N = \int^{\phi_*}_{\phi_{\rm end}} \frac{1+Q}{M_{\rm pl}^2}\frac{V}{V_{,\phi}}d\phi.
\end{align}
Once we determined the field value at the end of inflation $\phi_{\rm end}$, $N$ can be used to evaluate the field value at horizon crossing $\phi_*$. The value of $\phi_{\rm end}$ can be determined using the condition in Eq. (\ref{Eq:condition_of_end_of_inflation}) with Eq. (\ref{Eq:Q_V}) for a given value of $f$ and for the explicit form of potential $V$. To resolve the flatness problem, the required minimum number of e-folds is $N \simeq 40 - 60$.

\section{Analytic expressions of $n_{\mathrm{s}}$ and $\alpha_{\mathrm{s}}$ \label{section:exact_solutions}}
\subsection{Linear  dissipative coefficient}
The first objective of this study is to show a general analytical expression of $\alpha_{\mathrm{s}}$ without approximation for two major dissipative coefficients (linear and cubic). Therefore, we consider the cases $p=1$ \cite{Gil2016PRL} and $p=3$ \cite{Berera2009RPP,Gil2013JCAP}.

For $p=1$, we have the linear dissipative coefficient \cite{Gil2016PRL,Montefalcone2023JCAP}
\begin{equation}
\Gamma_{\rm linear}  = C T.
\end{equation}
In this case, the following function $G$ is obtained through numerical calculations \cite{Gil2011JCAP,Graham2009JCAP,Ramos2013JCAP,Gil2014JCAP,Gil2016PRL,Benetti2017PRD}:
\begin{equation}
G(Q) = 1 + 0.335 Q^{1.364} + 0.0185 Q^{2.315}.
\label{Eq:GQ_linear}
\end{equation}

The general exact expressions of the scalar spectral index $n_{\mathrm{s}}$ and the running scalar spectral index $\alpha_{\mathrm{s}}$ for the linear dissipative coefficient are obtained as follows:
\begin{align}
n_{\mathrm{s}} -1 = -\frac{18}{3+5Q}\epsilon + \frac{6}{3+5Q}\eta + \frac{2}{3+5Q}\left( \frac{Q}{G}\frac{dG}{dQ}  + 2AB \right)(3\epsilon-\eta)
\end{align}
and
\begin{align}
\alpha_{\mathrm{s}} &= -\frac{6(3\epsilon_{,N} - \eta_{,N})}{3+5Q} + \frac{30(3\epsilon - \eta)Q}{(3+5Q)^2} \frac{d\ln Q}{dN}\nonumber \\
   & \quad + \frac{2(3\epsilon - \eta)Q}{(3+5Q)G} \left\{ \frac{d}{d N}\frac{dG}{dQ}  + \frac{dG}{dQ} \left(\frac{3}{3+5Q} \frac{d\ln Q}{dN}-\frac{d\ln G}{dN} + \frac{3\epsilon_{,N} - \eta_{,N}}{3\epsilon-\eta} \right) \right\} \nonumber \\
    &\quad + \frac{4(3\epsilon-\eta)AQ}{3+5Q} \left\{\frac{B}{Q}\frac{d\ln A}{dN}- \frac{5 B}{3+5Q} \frac{d\ln Q}{dN}-  \frac{36\sqrt{3}\pi^2 Q}{(3+4\pi Q)^{5/2}}\frac{T}{H} (1+\pi Q)\frac{d\ln Q}{dN} \right. \nonumber \\
    &\quad  + \frac{6\sqrt{3}\pi }{(3+4\pi Q)^{3/2}}\frac{T}{H} \left[(1+2\pi Q)\frac{d\ln Q}{dN} + (1+\pi Q) \left( \frac{d\ln T}{dN} - \frac{d\ln H}{dN}\right) \right] \nonumber \\
    &\quad  + \left.  \frac{1}{Q}n_{\rm BE}^2e^{H/T} \frac{H}{T} \left[2\cdot\frac{d \ln n_{\rm BE}}{dN} + \left(\frac{H}{T}+1\right)\left(\frac{d\ln H}{dN} -\frac{d\ln T}{dN}\right) \right]\right\} \nonumber \\
    &\quad  +\frac{4(3\epsilon_{,N}-\eta_{,N})AB}{3+5Q},
\end{align}
respectively. As expected, in the limits $Q \rightarrow 0$ and $T \rightarrow 0$, the following relations in the CI are recovered:
\begin{align}
n_{\mathrm{s}} -1 = -6\epsilon + 2\eta, \quad \alpha_{\mathrm{s}} = -24\epsilon^2 + 16\epsilon \eta -2\xi_2,
\end{align}
where
\begin{align}
\xi_2 = \epsilon\xi=M_{\rm pl}^4 \frac{V_{,\phi} V_{,\phi\phi\phi} }{V^2}.
\end{align}
%

\subsection{Cubic dissipative coefficient}
For $p=3$ in Eq.(\ref{Eq:GammaTf}), we have the cubic dissipative coefficient \cite{Berera2009RPP,Gil2011JCAP,Gil2013JCAP,Gil2009IJMPA,Montefalcone2023JCAP}
\begin{equation}
\Gamma_{\rm cubic}  = C \frac{T^3}{f_1^2}
\end{equation}
with the function $G$ \cite{Gil2011JCAP,Graham2009JCAP,Ramos2013JCAP,Gil2014JCAP,Gil2016PRL,Benetti2017PRD}:
\begin{equation}
G(Q) = 1 + 4.981 Q^{1.946} + 0.127 Q^{4.330}.
\label{Eq:GQ_cubic}
\end{equation}

The general exact expressions of the scalar spectral index $n_{\mathrm{s}}$ and the running scalar spectral index $\alpha_{\mathrm{s}}$ for the cubic dissipative coefficient are obtained as follows:
\begin{align}
n_{\mathrm{s}} -1 =& -\frac{2(3+11Q)}{(1+Q)(1+7Q)}\epsilon + \frac{2}{1+7Q}\eta + \frac{16Q}{(1+Q)(1+7Q)}\sigma  \nonumber \\
& + \frac{2}{1+7Q}\frac{Q}{G}\frac{dG}{dQ}(5\epsilon-3\eta+4\sigma) + \frac{AD}{(1+Q)(1+7Q)}
\end{align}
and
\begin{align}
\alpha_{\mathrm{s}} &= -\frac{2(3+11Q)\epsilon_{,N}-2(1+Q)\eta_{,N} - 16Q\sigma_{,N}}{(1+Q)(1+7Q)} \nonumber \\
& \quad + \frac{ 2\left[ (13+42Q+77Q^2)\epsilon - 7(1+Q)^2\eta+8(1-7Q^2)\sigma \right]Q}{(1+Q)^2(1+7Q)^2} \frac{d\ln Q}{dN} \nonumber \\
& \quad + \frac{2(5\epsilon - 3\eta +4\sigma)Q}{(1+7Q)G} \left\{ \frac{d}{d N}\frac{dG}{dQ}  + \frac{dG}{dQ} \left(\frac{1}{1+7Q} \frac{d\ln Q}{dN}-\frac{d\ln G}{dN}  \right.\right. \nonumber \\
& \quad \left.\left. + \frac{5\epsilon_{,N} - 3\eta_{,N} + 4\sigma_{,N}}{5\epsilon - 3\eta +4\sigma} \right) \right\} \nonumber \\
& \quad + \frac{AQ}{(1+Q)(1+7Q)}\left\{ 4n_{\rm BE}^2e^{H/T}  \frac{1}{Q}\frac{H}{T} \left[ 2(1+2Q)\epsilon_{,N} - (1+Q)\eta_{,N} +(1-Q) \sigma_{,N} \right]\right. \nonumber \\
& \quad + \frac{4\sqrt{3}\pi}{(3+4\pi Q)^{3/2}}\frac{T}{H} \left[ (21+9(3+2\pi)Q+26\pi Q^2)\epsilon_{,N} -2(1+Q)(6+5\pi Q)\eta_{,N}  \right. \nonumber \\
& \quad +\left. \left. (15 + 3(3+4\pi)Q+4\pi Q^2)\sigma_{,N} \right] \right\}\nonumber \\
& \quad +  \frac{AD}{(1+Q)(1+7Q)} \frac{d\ln A}{dN}+ \frac{AQ}{(1+Q)(1+7Q)} \left\{4n_{\rm BE}^2e^{H/T} \left[ 2(1+2Q)\epsilon \right. \right. \nonumber \\
 & \qquad \left. - (1+Q)\eta +(1-Q) \sigma \right] \times  \left. \frac{1}{Q}\frac{H}{T} \left[ 2\cdot\frac{d\ln n_{\rm BE}}{dN} + \left(1+\frac{H}{T}\right) \left(\frac{d\ln H}{dN} - \frac{d\ln T}{dN} \right) \right] \right. \nonumber \\
& \quad - \left. \frac{1}{(1+Q)(1+7Q)}\frac{H}{T} \left[  4n_{\rm BE}^2e^{H/T} \left(4(3+7Q+7Q^2)\epsilon -7(1+Q)^2\eta \right. \right.\right. \nonumber \\
& \qquad + \left.  \left.  (9+14Q-7Q^2)\sigma  \right) \right]\frac{d\ln Q}{dN}   \nonumber \\
& \quad + \frac{4\sqrt{3}\pi}{(3+4\pi Q)^{3/2}}\frac{T}{H} \left[ (21+9(3+2\pi)Q+26\pi Q^2)\epsilon - 2(1+Q)(6+5\pi Q)\eta \right. \nonumber \\
& \qquad + \left. (15+3(3+4\pi)Q + 4\pi Q^2)\sigma \right]\left( \frac{d\ln T}{dN} - \frac{d\ln H}{dN} \right)   \nonumber \\
& \quad +  \frac{4\sqrt{3}\pi}{(3+4\pi Q)^{3/2}}\frac{T}{H} \left[ Q\left( (27+18\pi+52\pi Q)\epsilon - 2(6+5\pi+10\pi Q)\eta \right.\right. \nonumber \\
& \quad \left. + (9+12\pi+8\pi Q)\sigma \right) \nonumber \\
& \quad + \left. \frac{3-2\pi Q-3(7+16\pi)Q^2-70\pi Q^3}{(1+Q)(1+7Q)(3+4\pi Q)} \left( (21+9(3+2\pi)Q+26\pi Q^2)\epsilon \right.\right.\nonumber \\
& \quad - \left. \left.  \left.  2(1+Q)(6+5\pi Q)\eta + (15 + 3(3+4\pi)Q+4\pi Q^2)\sigma \right) \right] \frac{d\ln Q}{dN} \right\},
\end{align}
respectively. As same as the case of the linear dissipative coefficient, the relations in the CI are recovered in the limits $Q \rightarrow 0$ and $T \rightarrow 0$.

\subsection{Comparison with a previous study }
\begin{figure}[t]
\centering
\includegraphics[keepaspectratio, scale=0.40]{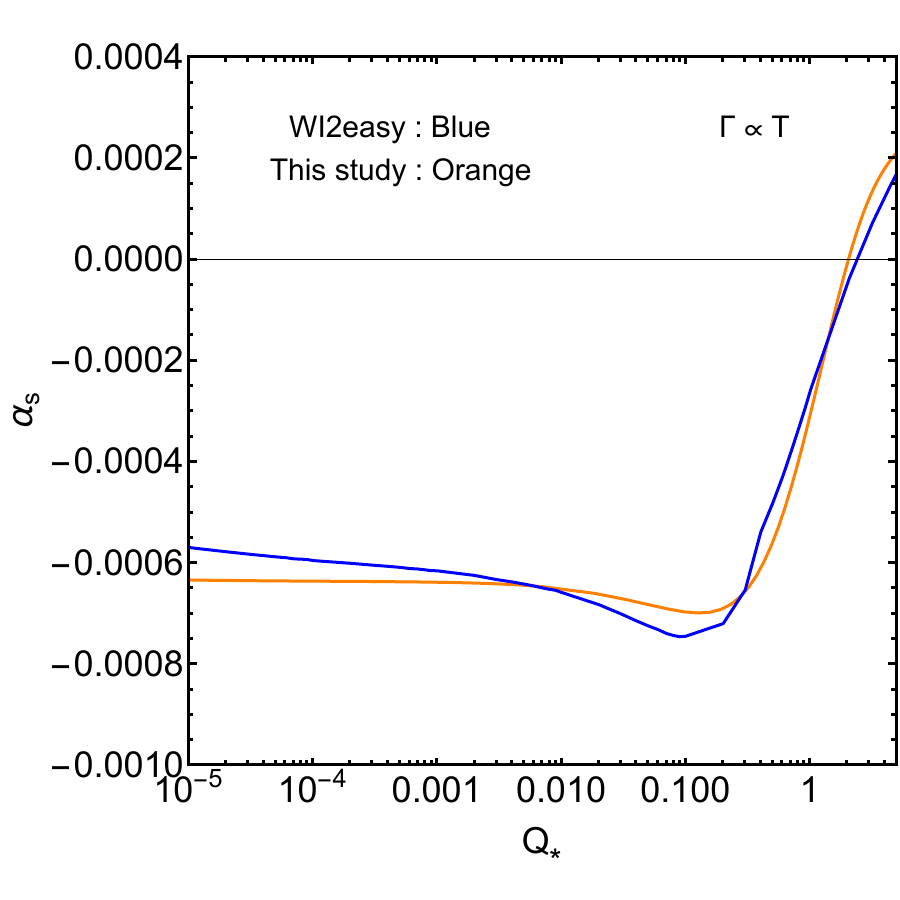}
\hspace{2mm}
\includegraphics[keepaspectratio, scale=0.40]{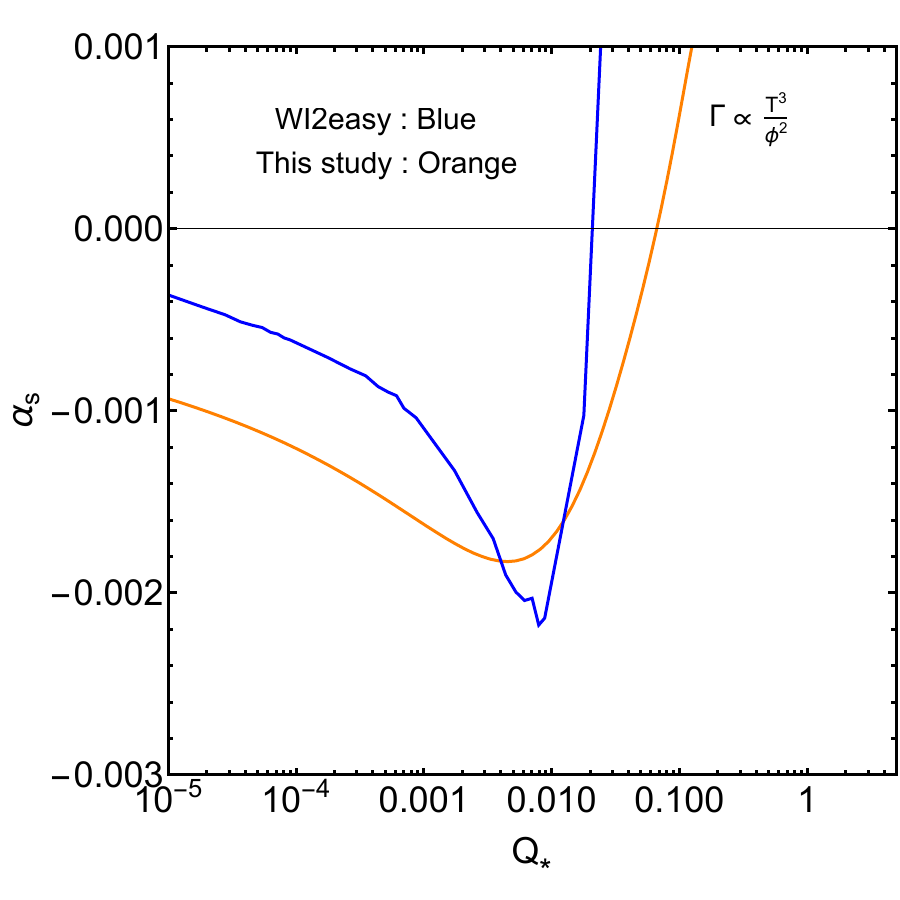}
 \caption{Comparison with a previous study.}
 \label{FIG:Comparision}
 \end{figure}

We demonstrate that the analytical expression of $\alpha_s$ obtained in this study is consistent with the results from numerical calculation by WI2easy  \cite{Rodrigues2025arXiv}. Figure \ref{FIG:Comparision} shows the predictions of $\alpha_s$ in the warm chaotic inflation potential $V=\lambda \phi^4/4$. The left (right) panel shows  $\alpha_s$ vs $Q_*$, dimensionless dissipative ratio at the Hubble radius crossing, for the linear (cubic) dissipative coefficient. In the both panels, the blue curve depicts the results from WI2easy and the orange curve shows the prediction using by the analytical expressions in this study. Since we use approximate expressions of $G(Q)$ in Eqs. (\ref{Eq:GQ_linear}) or (\ref{Eq:GQ_cubic}) for linear or cubic dissipative coefficient but WI2easy always use the numerically derived data for $G(Q)$, a little differences are observed in the blue and orange curves. We can conclude that the analytical expression of $\alpha_s$ is consistent with the WI2easy. This observation shows the availability of not only the analytical expression of $\alpha_s$ but also the WI2easy code.

\section{Warm natural inflation\label{section:WNI}}

\subsection{Slow-roll parameters and other key relations}
The second goal of this study is to evaluate the running scalar index $\alpha_{\mathrm{s}}$ in the WNI model. The inflaton potential in the WNI and CNI models is expressed as
\begin{equation}
V(\phi) = \Lambda^4 \left[ 1 + \cos \left( \frac{\phi}{f} \right) \right],
\end{equation}
where $\phi$ denotes the inflaton (an axion-like particle), $\Lambda$ denotes the inflation scale, and $f$ represents the decay constant, which is related to the symmetry-breaking scale associated with the spontaneous symmetry breaking of the global symmetry of the inflaton. The slow-roll parameters and other key relations for the WNI model are obtained as follows:
\begin{equation}
\epsilon = \frac{1}{2\tilde{f}^2} \frac{\sin^2 (\tilde{\phi}/\tilde{f})}{\left[1 + \cos (\tilde{\phi}/\tilde{f}) \right]^2},
\end{equation}
\begin{equation}
\eta = - \frac{1}{\tilde{f}^2} \frac{\cos (\tilde{\phi}/\tilde{f})}{1 + \cos (\tilde{\phi}/\tilde{f})},
\end{equation}
\begin{equation}
\xi = -  \frac{2}{\tilde{f}^2},
\end{equation}
and \footnote{Eq.(3.7) in \cite{Montefalcone2023JCAP} may be corrected as Eq.(\ref{Eq:TH}).}
\begin{equation}
\frac{T}{H} = \left\{ \frac{9}{4\tilde{g}_*}\frac{Q}{(1+Q)^2}\frac{M_{\rm pl}^4}{\tilde{f}^2\Lambda^4}  \frac{\sin^2(\tilde{\phi}/\tilde{f})}{\left[1 + \cos (\tilde{\phi}/\tilde{f})\right]^3 } \right\}^{1/4},
\label{Eq:TH}
\end{equation}
where we have defined $\tilde{f} = f / M_{\rm pl}$ and $\tilde{\phi} = \phi / M_{\rm pl}$. Note that $\sigma = 0$ in this study. The number of e-folds can be estimated as follows:
\begin{align}
N = \int_{\tilde{\phi}_*}^{\tilde{\phi}_{\rm end}} (1+Q) \tilde{f} \frac{1+\cos(\tilde{\phi}/\tilde{f})}{\sin(\tilde{\phi}/\tilde{f})}d\tilde{\phi}.
\end{align}

The dimensionless dissipative ratio $Q$ can be determined using the following relations \footnote{Eq.(3.8) in \cite{Montefalcone2023JCAP} may be corrected as Eqs. (\ref{Eq:WNI_Q_linear}) and (\ref{Eq:WNI_Q_cubic}).}:
\begin{align}
Q^{3}(1+Q)^{2} = \frac{c_1}{\tilde{f}^2} \frac{\sin^2(\tilde{\phi}/\tilde{f})}{\left[1+\cos (\tilde{\phi}/\tilde{f})\right]^3 },
\label{Eq:WNI_Q_linear}
\end{align}
with
\begin{align}
c_1= \frac{C^4 M_{\rm pl}^4}{36 \tilde{g}_* \Lambda^4}
\label{Eq:c1}
\end{align}
for the linear dissipative coefficient $\Gamma_{\rm linear}  = C T$, and
\begin{align}
Q(1+Q)^{6} = \frac{c_3}{\tilde{f}^{6}} \frac{\sin^6(\tilde{\phi}/\tilde{f})}{\left[1+\cos (\tilde{\phi}/\tilde{f})\right]^5 },
\label{Eq:WNI_Q_cubic}
\end{align}
with
\begin{align}
c_3 = \frac{C^4 M_{\rm pl}^4 \Lambda^4}{576 \tilde{g}_*^3 f_1^8}
\label{Eq:c3}
\end{align}
for the cubic dissipative coefficient $\Gamma_{\rm cubic} = C T^3/f_1^2$.
\begin{figure}[t]
\centering
\includegraphics[keepaspectratio, scale=0.4]{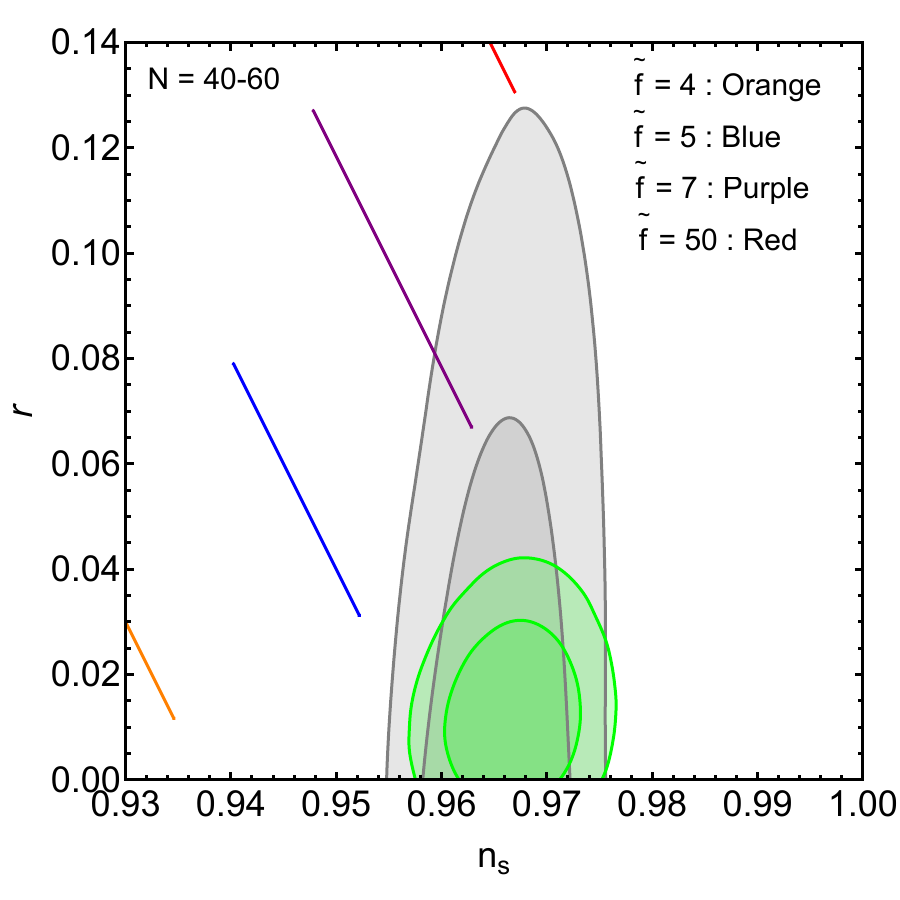}
\includegraphics[keepaspectratio, scale=0.4]{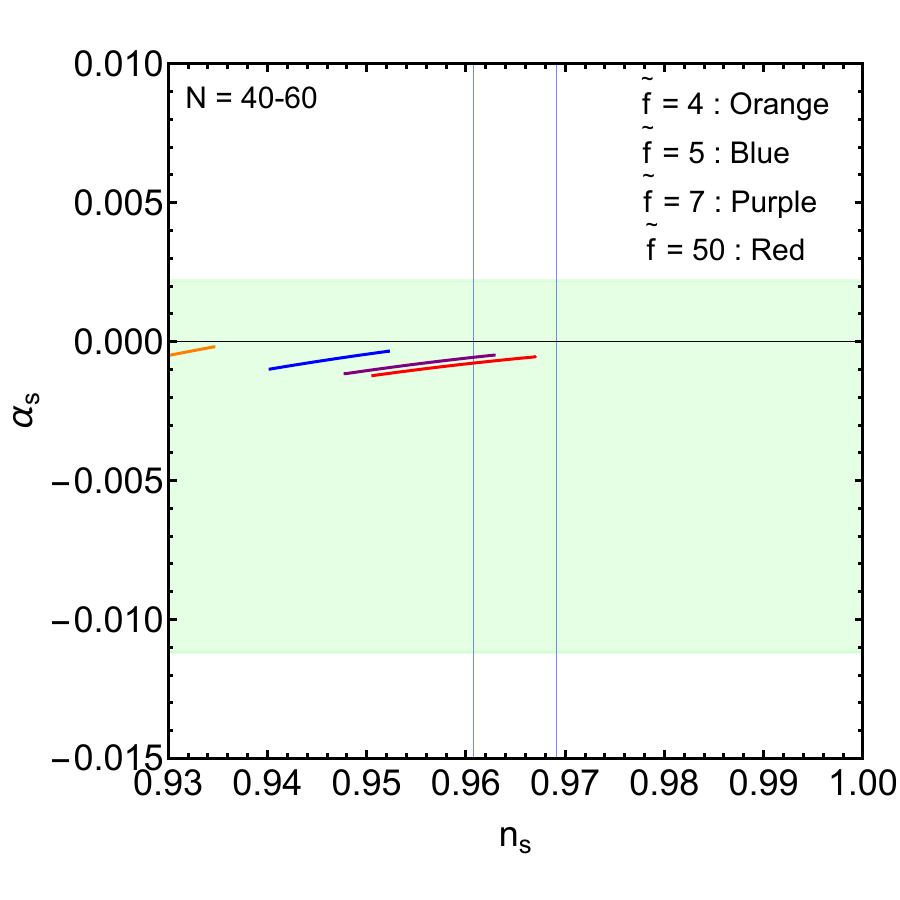}
 \caption{$n_{\mathrm{s}}$-$r$ (left) and $n_{\mathrm{s}}$--$\alpha_{\mathrm{s}}$ (right) in CNI ($Q=0$ case).}
 \label{FIG:CNI}
 \end{figure}

\begin{figure}[h]
\centering
\includegraphics[keepaspectratio, scale=0.27]{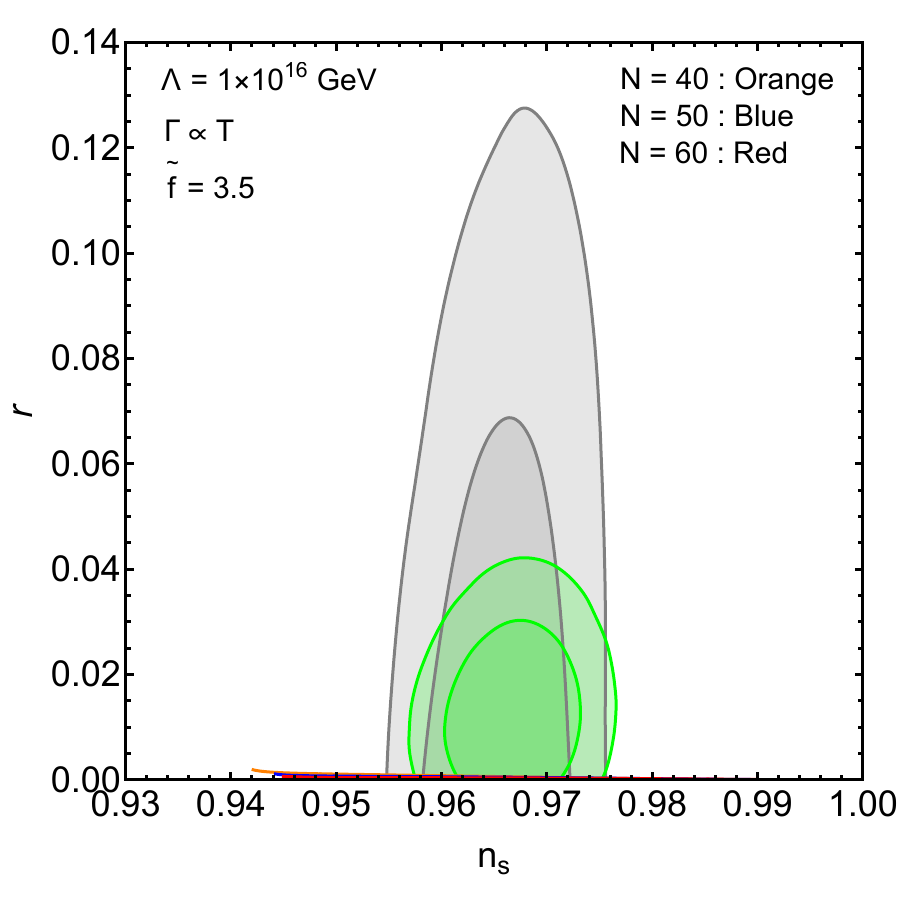}
\includegraphics[keepaspectratio, scale=0.27]{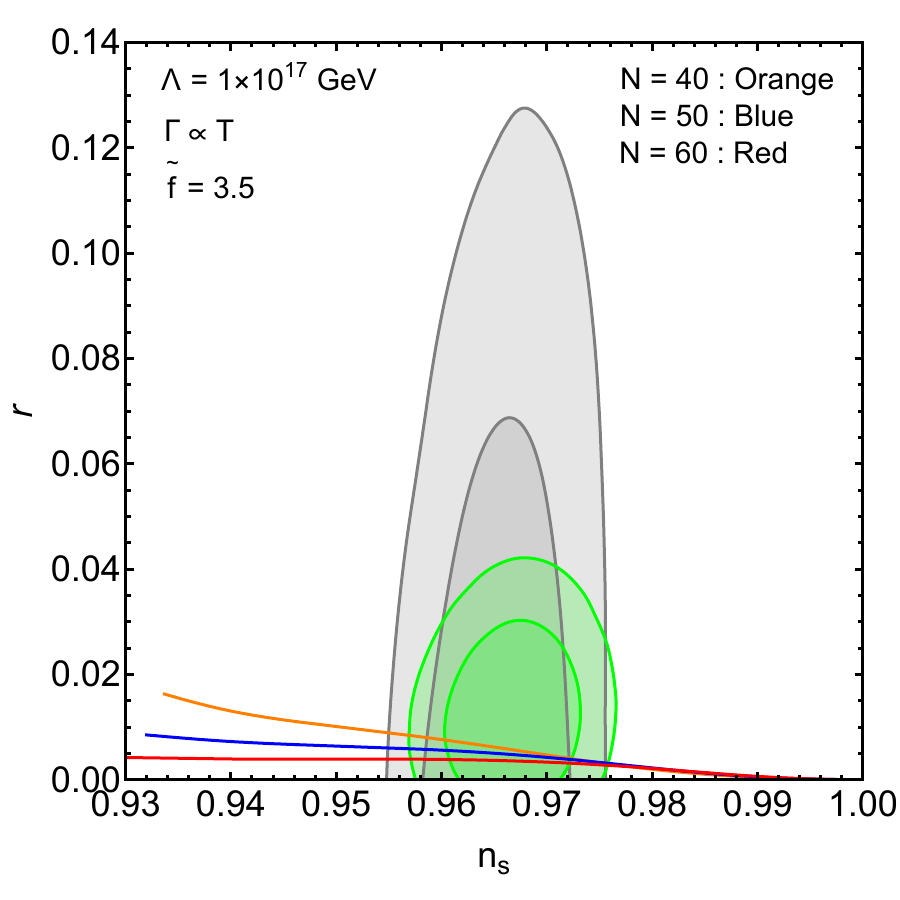}
\includegraphics[keepaspectratio, scale=0.27]{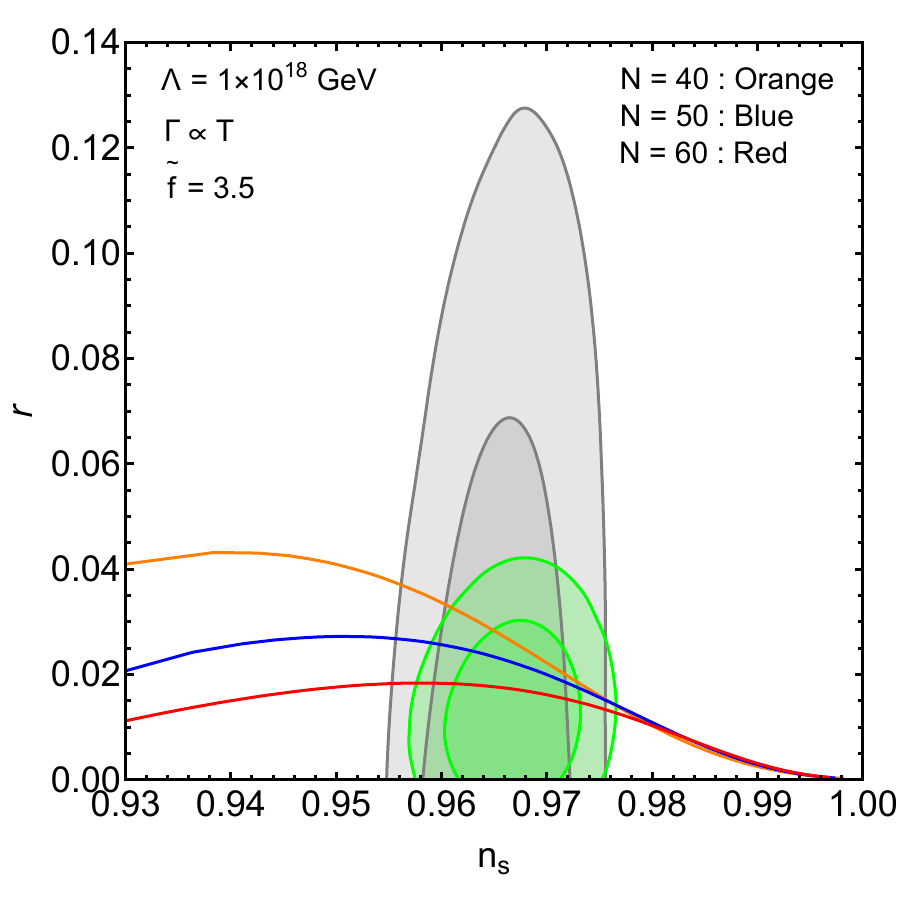}\\
\includegraphics[keepaspectratio, scale=0.27]{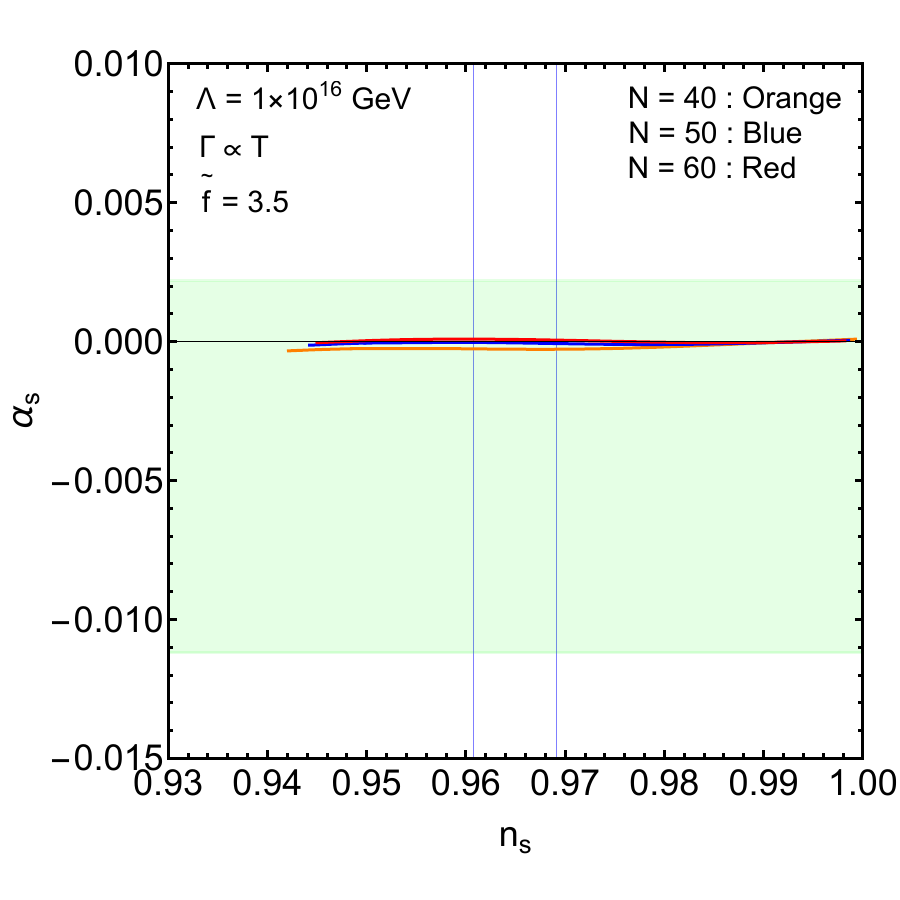}
\includegraphics[keepaspectratio, scale=0.27]{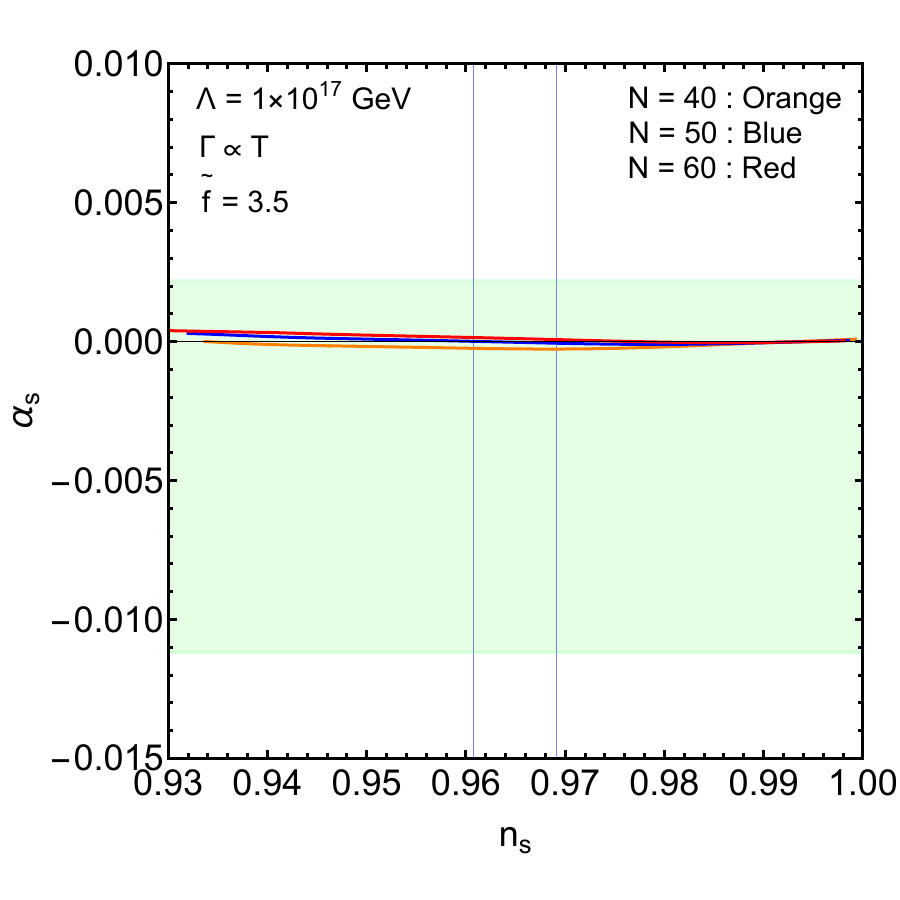}
\includegraphics[keepaspectratio, scale=0.27]{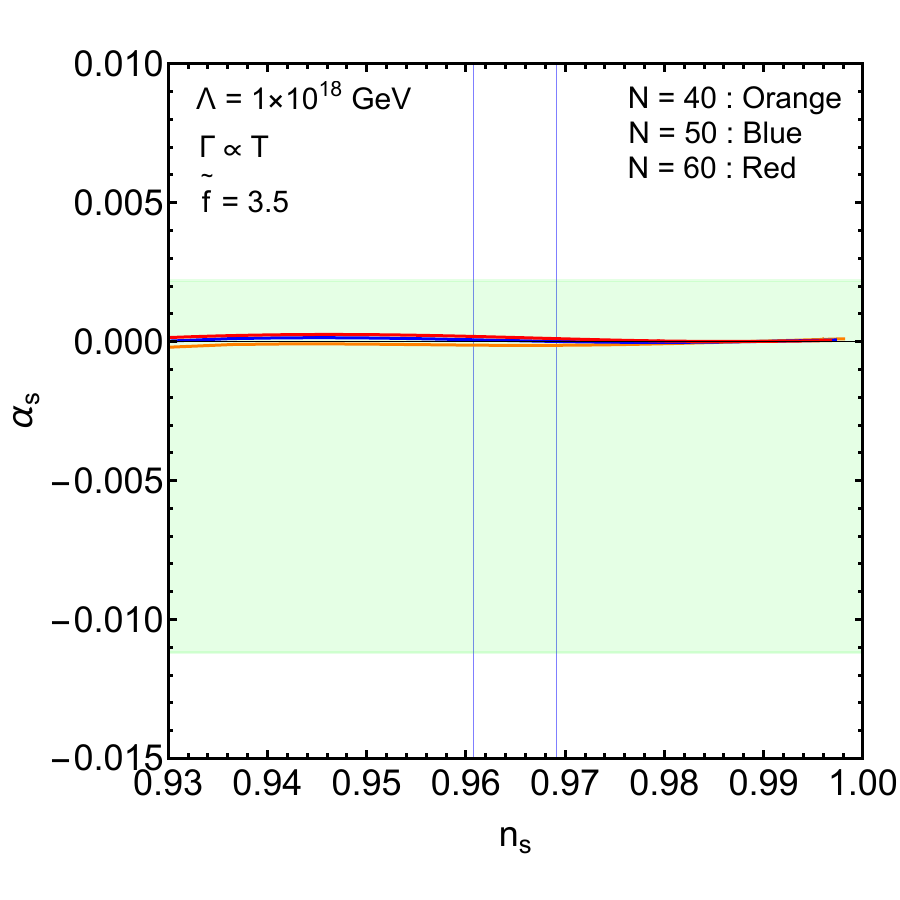}\\
\includegraphics[keepaspectratio, scale=0.27]{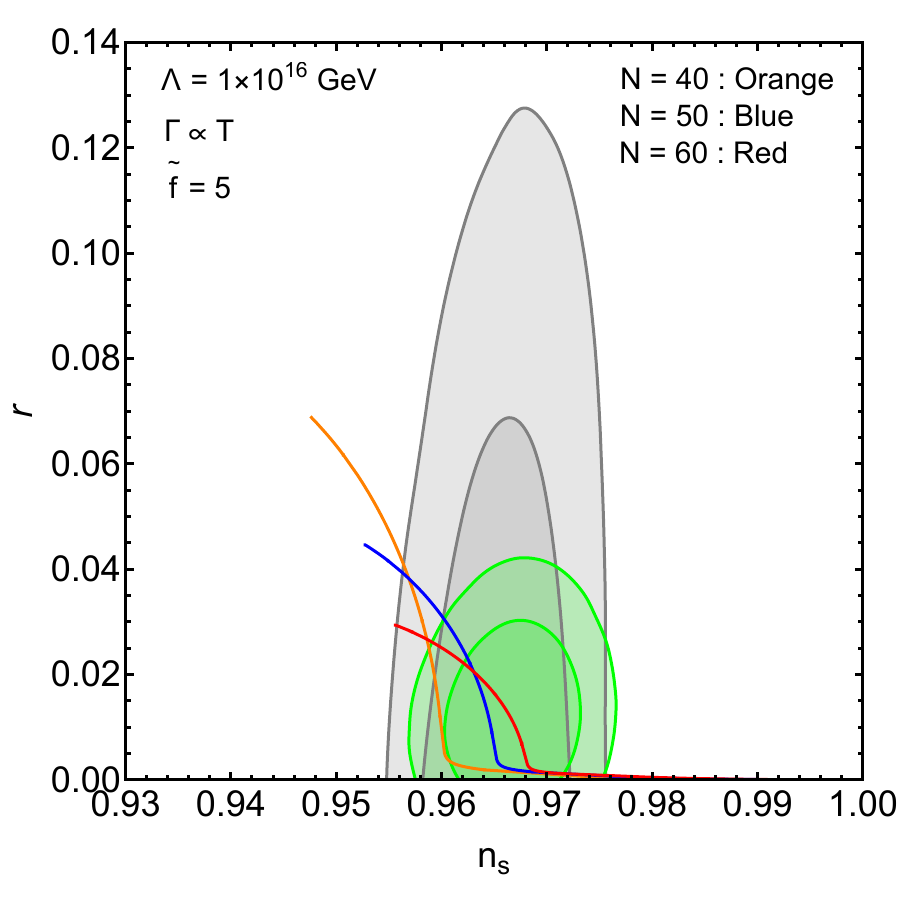}
\includegraphics[keepaspectratio, scale=0.27]{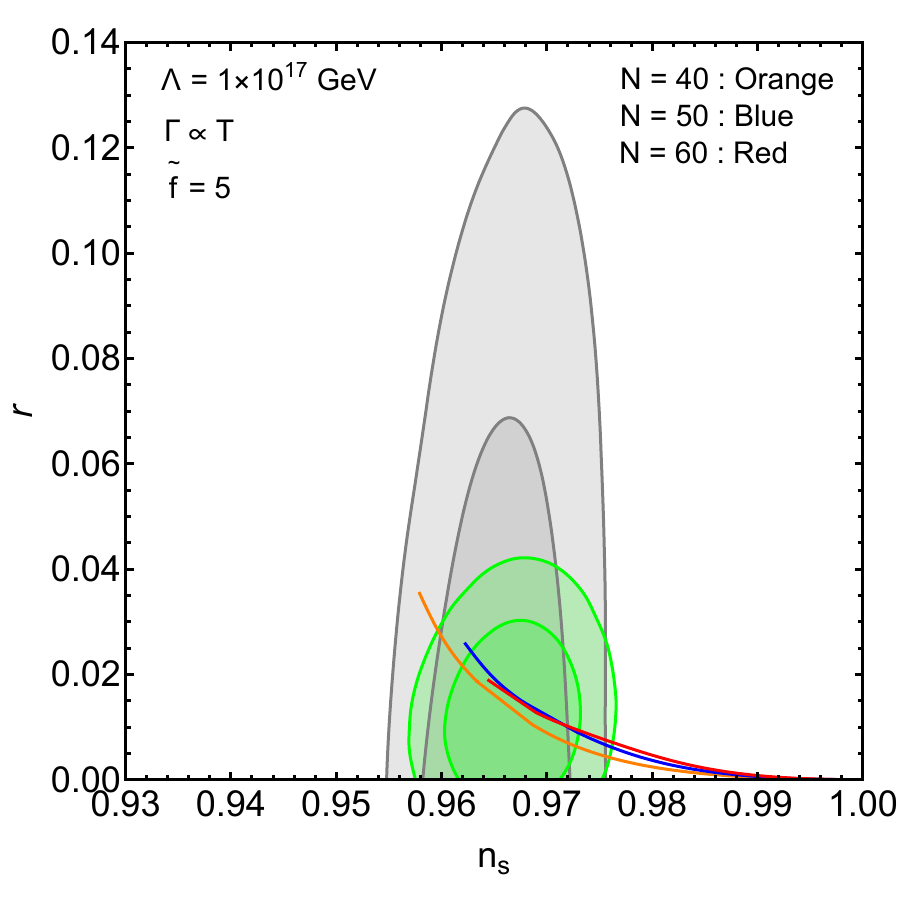}
\includegraphics[keepaspectratio, scale=0.27]{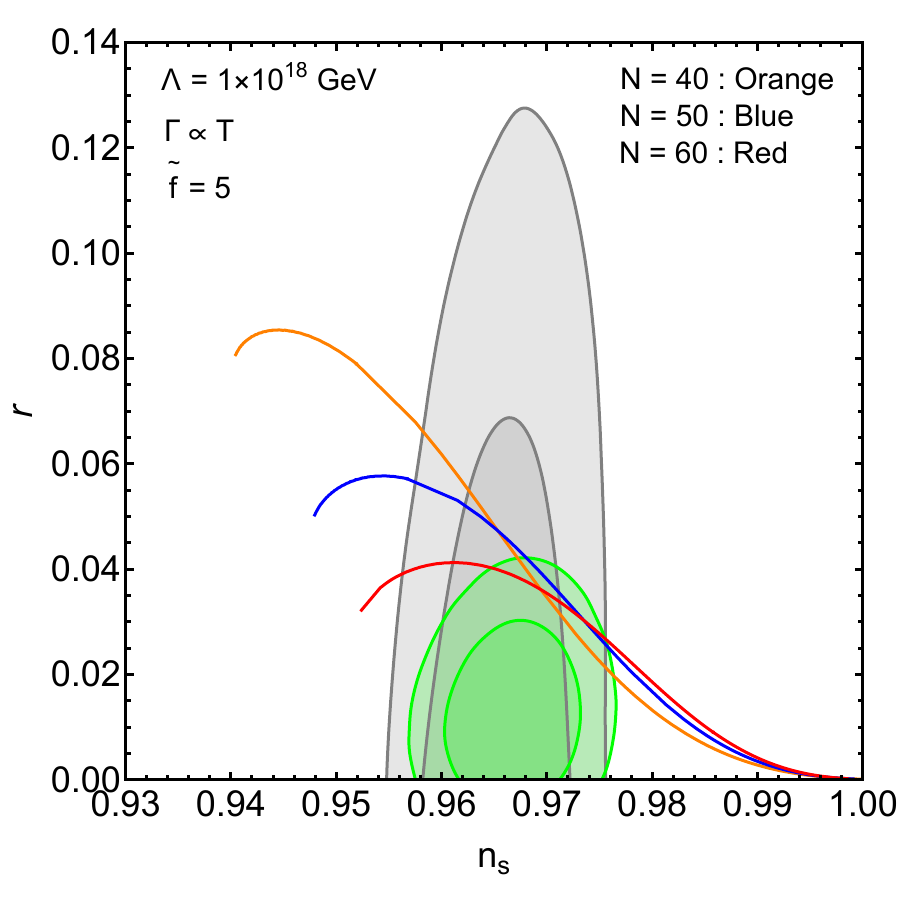}\\
\includegraphics[keepaspectratio, scale=0.27]{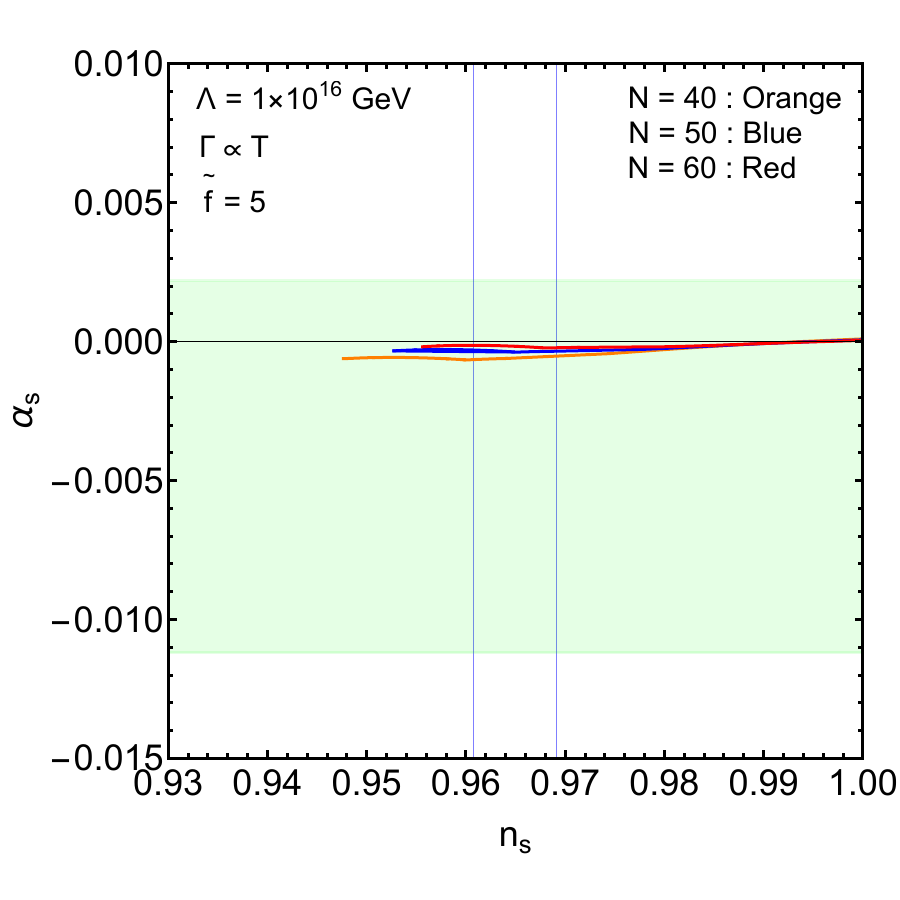}
\includegraphics[keepaspectratio, scale=0.27]{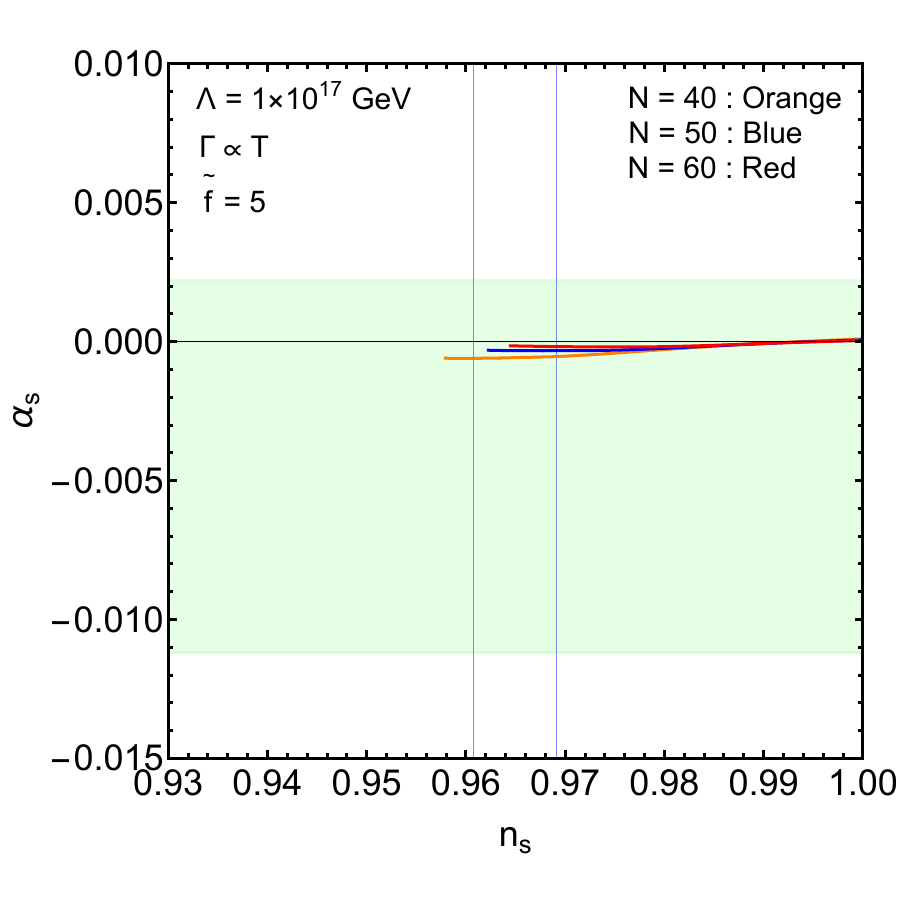}
\includegraphics[keepaspectratio, scale=0.27]{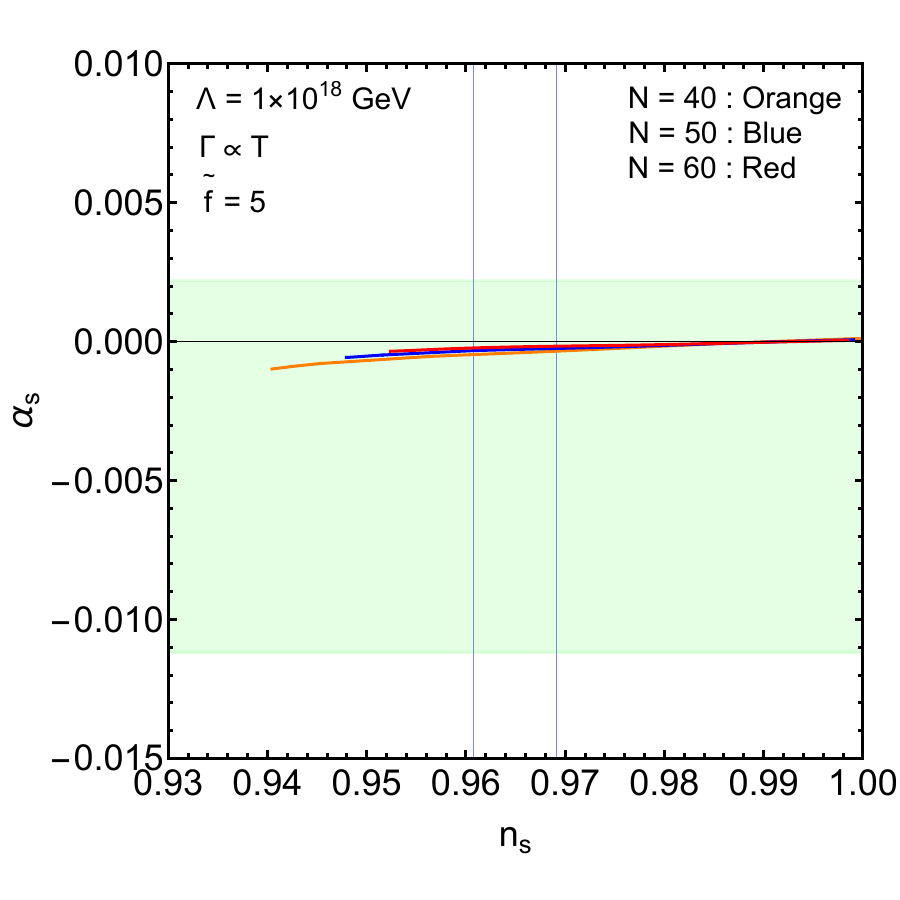}\\
 \caption{$n_{\mathrm{s}}$, $r$ and $\alpha_{\mathrm{s}}$ in the WNI with the linear dissipative coefficient.}
 \label{FIG:WNI_linear}
 \end{figure}

\begin{figure}[h]
\centering
\includegraphics[keepaspectratio, scale=0.27]{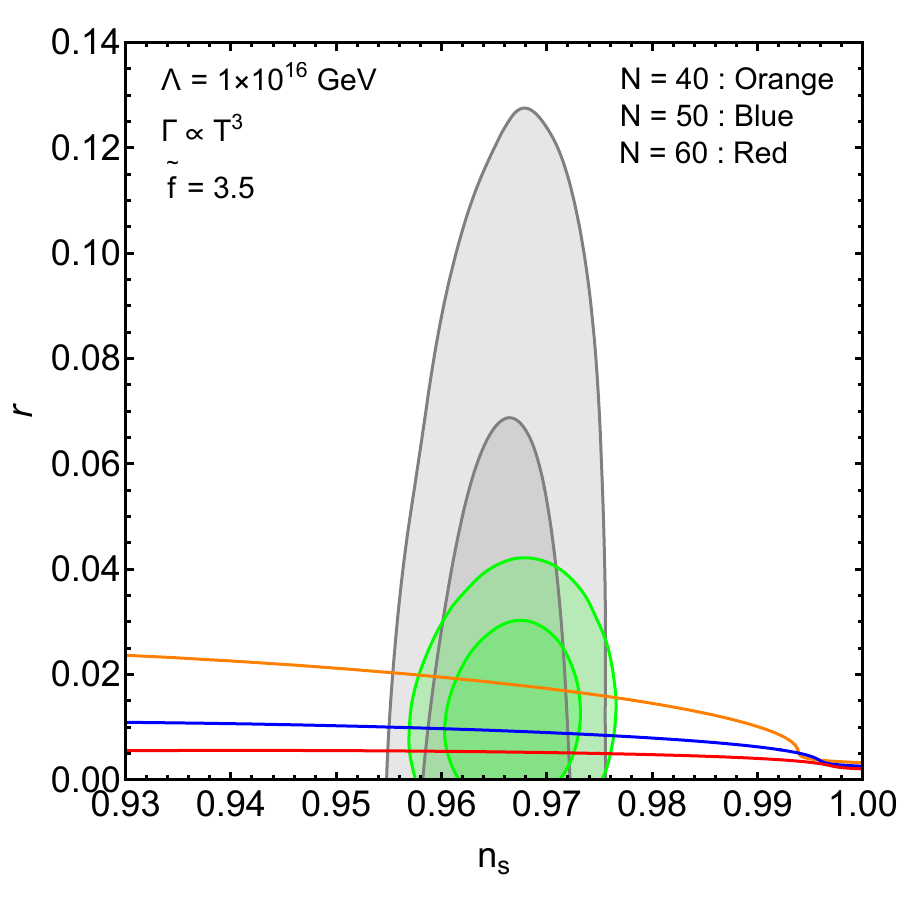}
\includegraphics[keepaspectratio, scale=0.27]{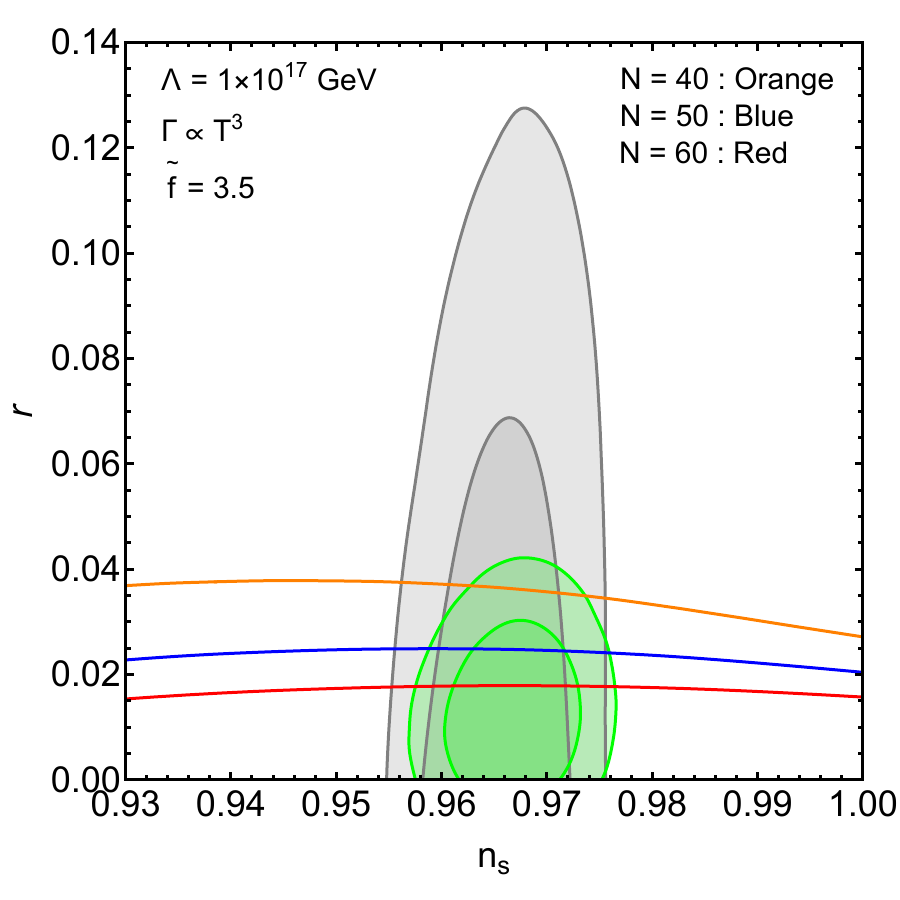}
\includegraphics[keepaspectratio, scale=0.27]{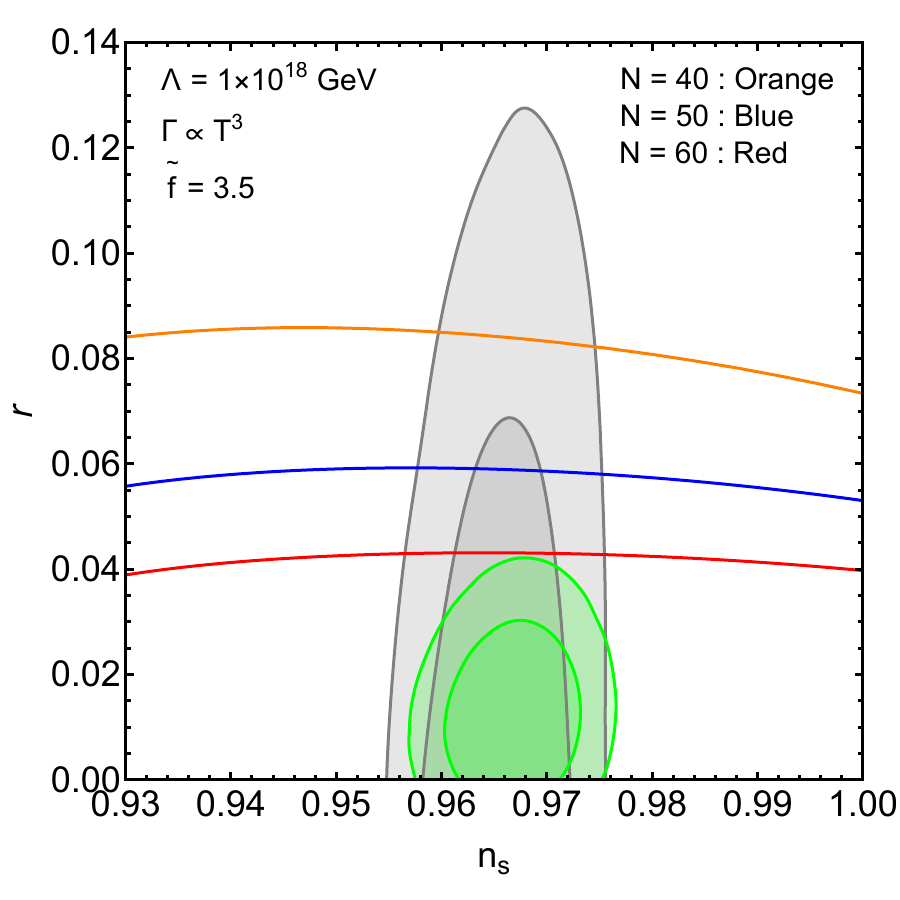}\\
\includegraphics[keepaspectratio, scale=0.27]{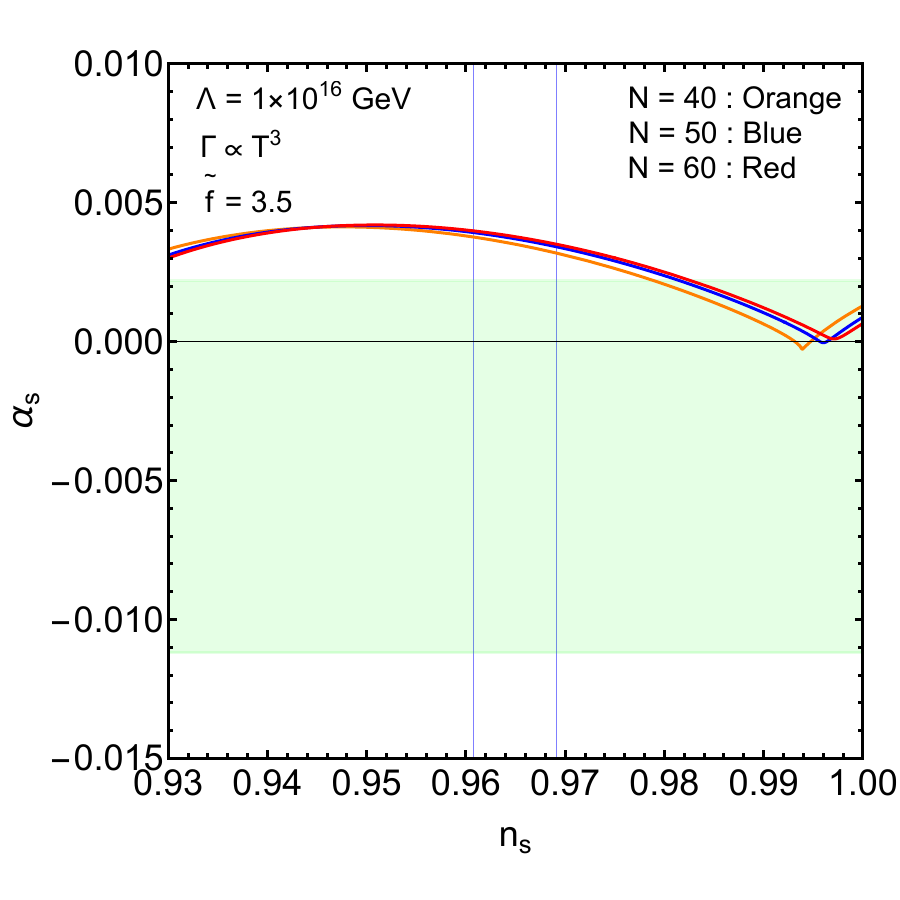}
\includegraphics[keepaspectratio, scale=0.27]{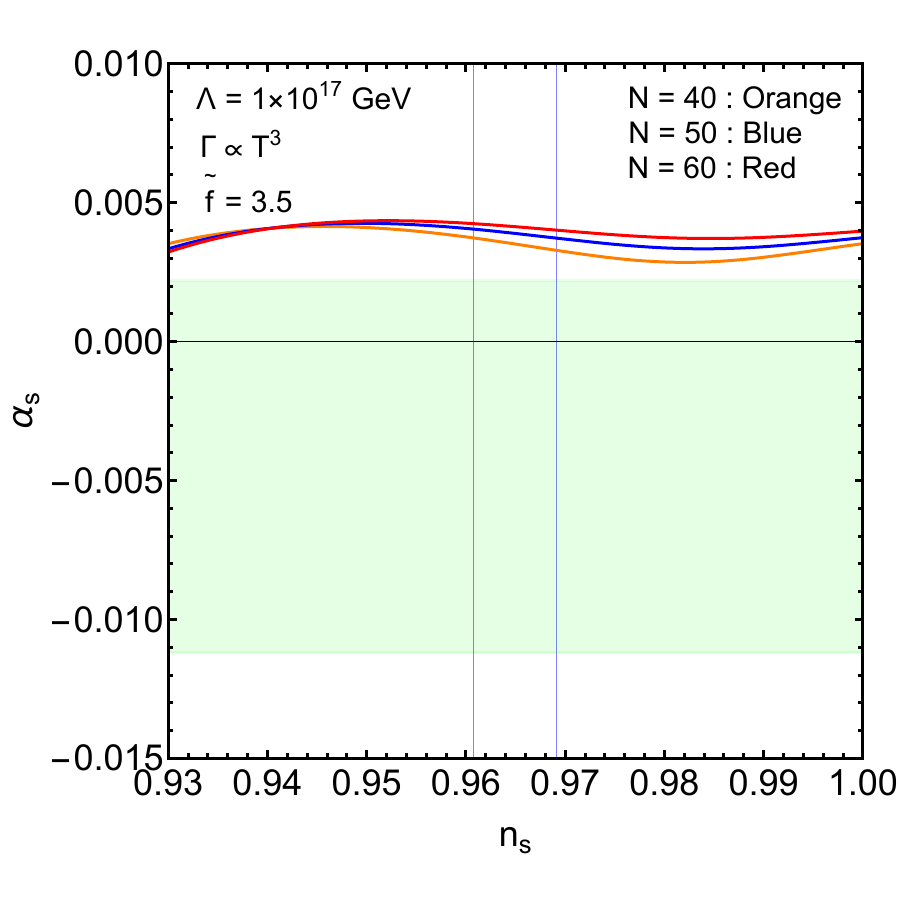}
\includegraphics[keepaspectratio, scale=0.27]{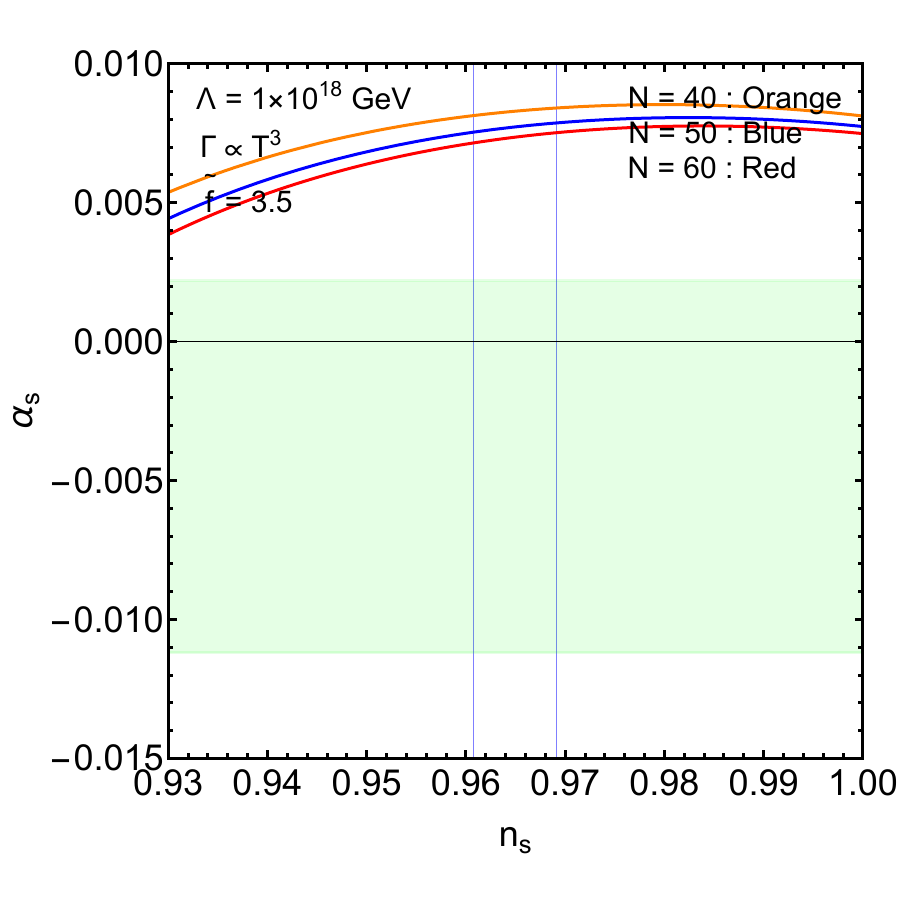}\\
\includegraphics[keepaspectratio, scale=0.27]{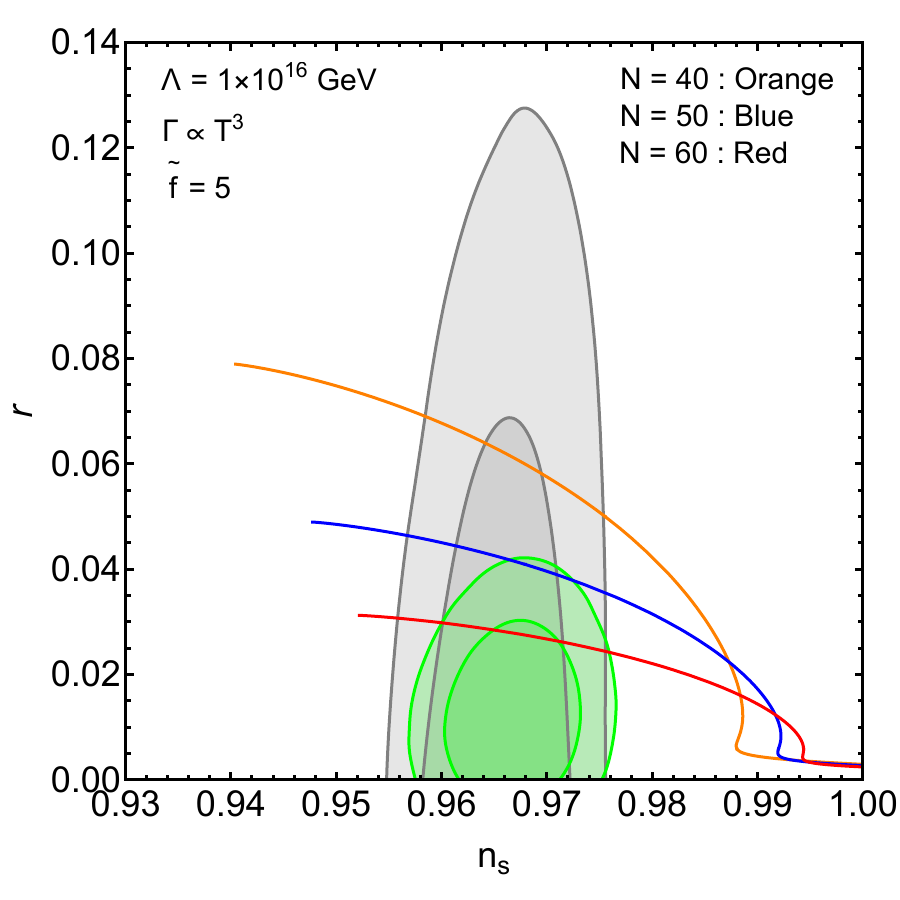}
\includegraphics[keepaspectratio, scale=0.27]{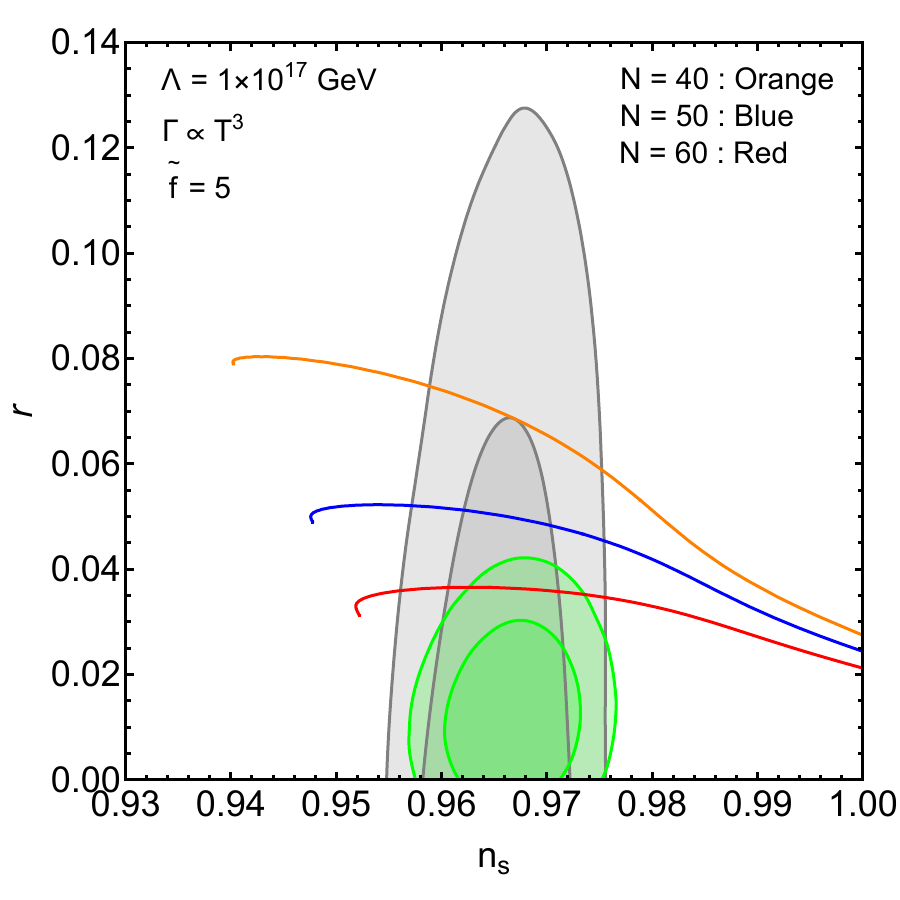}
\includegraphics[keepaspectratio, scale=0.27]{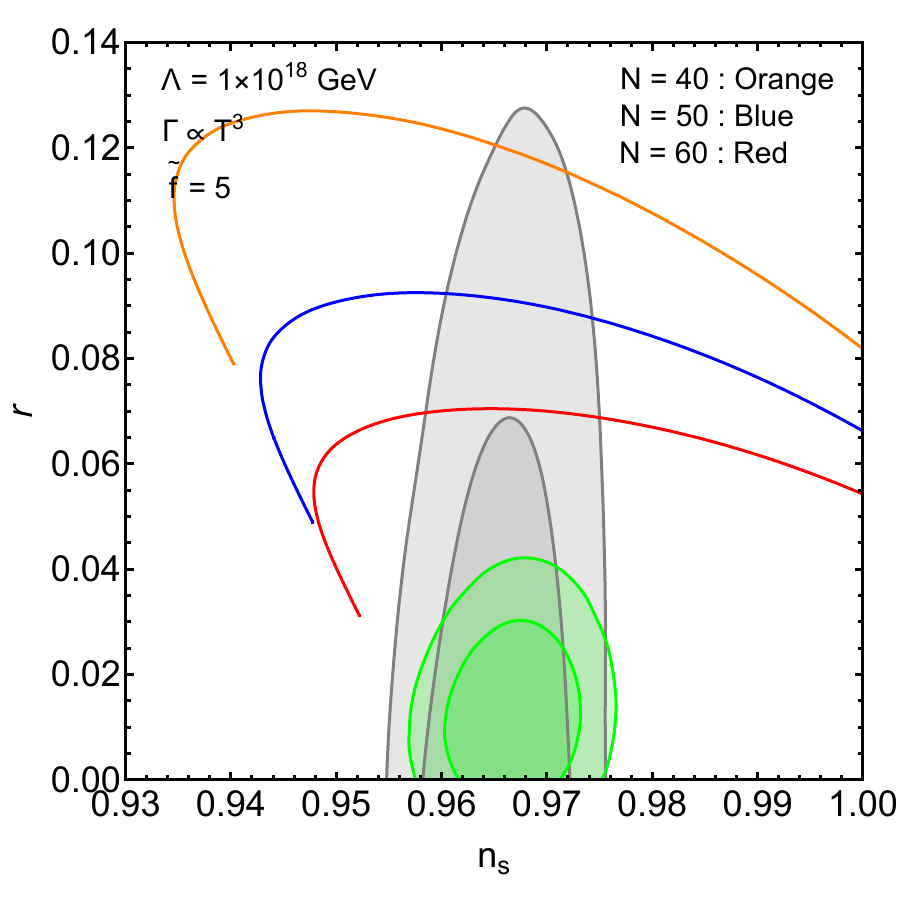}\\
\includegraphics[keepaspectratio, scale=0.27]{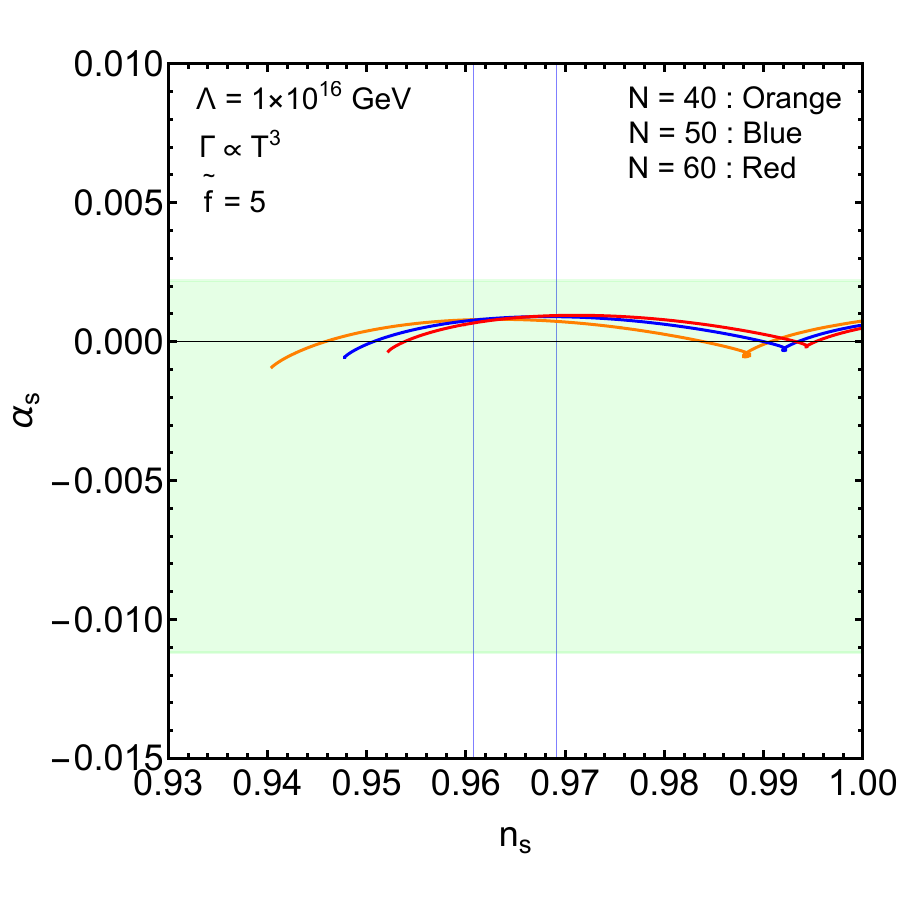}
\includegraphics[keepaspectratio, scale=0.27]{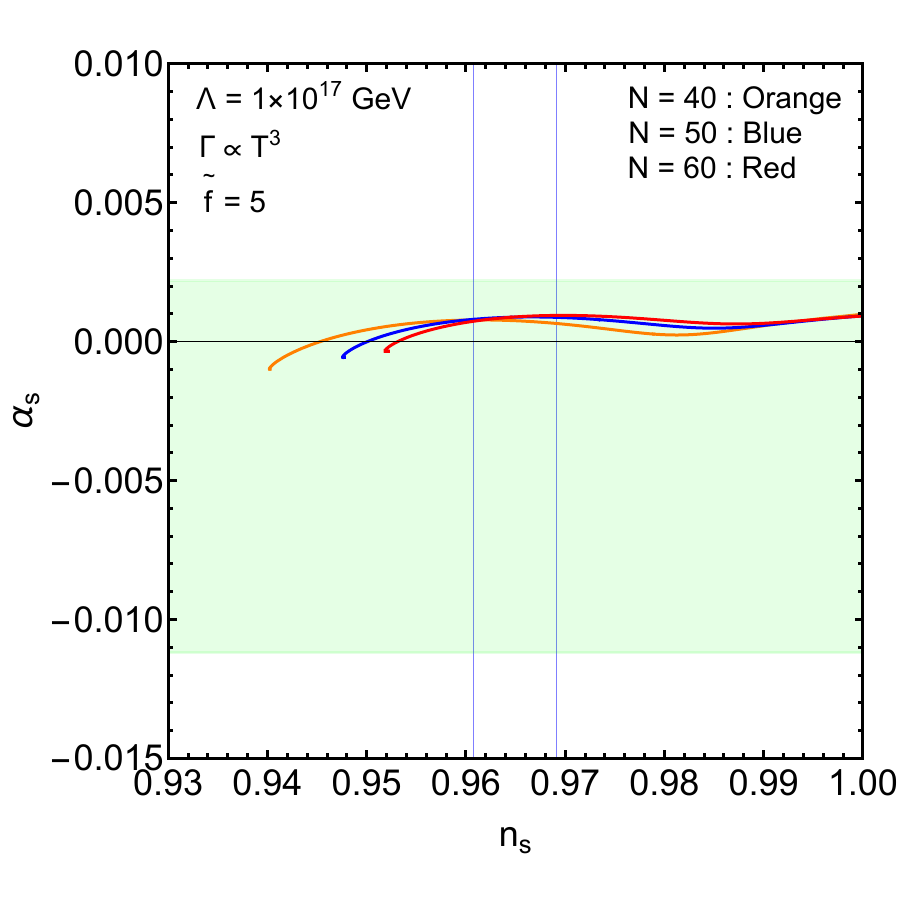}
\includegraphics[keepaspectratio, scale=0.27]{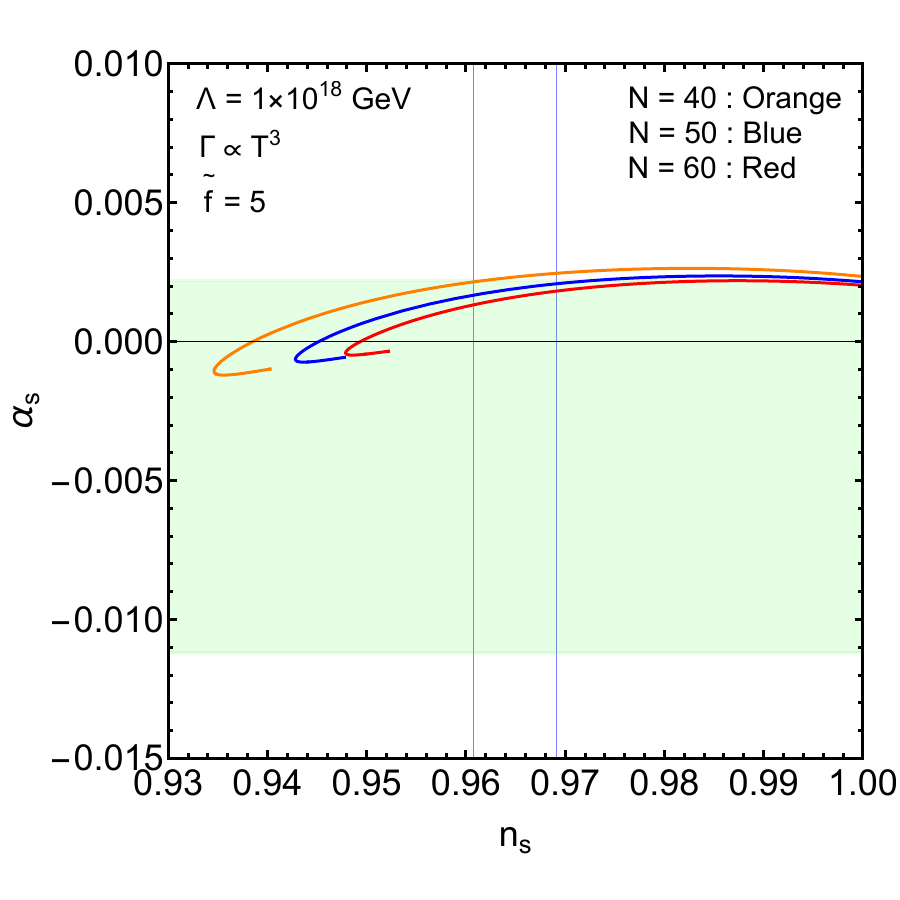}\\
 \caption{The same as Fig. \ref{FIG:WNI_linear} but with the cubic dissipative coefficient.}
 \label{FIG:WNI_cubic}
 \end{figure}

\subsection{Numerical analysis}
The validity of WI (as well as CI) models can be evaluated according to the consistency of the predicted scalar spectral index $n_{\mathrm{s}}$, the tensor scalar ratio $r$, and the running scalar spectral index $\alpha_{\mathrm{s}}$ with CMB observations. For the $\Lambda$CDM model, the Planck 2018 TT,TE,EE+lowE+lensing data constrain the scalar spectral index and its running at 68\% confidence level (CL) as follows \cite{PLANCK2020AA_VI}:
\begin{align}
n_{\mathrm{s}} &= 0.9649 \pm 0.0042, \\
\alpha_{\mathrm{s}} &= -0.0045 \pm 0.0067.
\end{align}
The tensor scalar ratio is constrained at 95\% CL through BICEP/Keck analysis \cite{BICEP22018PRL,BICEPKeck2021PRL}:
\begin{align}
r < 0.036.
\end{align}

Before presenting the results of the numerical analysis for WNI, we present the CNI results ($Q=0$ case) for $\tilde{f}=\{4, 5, 7, 50\}$ and $N=40$--$60$. In the left panel in Fig. \ref{FIG:CNI}, the light gray (light green) and gray (green) regions show the constraints of $n_{\mathrm{s}}$ and $r$ from Planck 2018 (Planck with BICEP/Keck) at 98\% and 68\% CL. The colored curves denote the prediction of $n_{\mathrm{s}}$ and $r$. The right panel shows the prediction of $n_{\mathrm{s}}$ and $\alpha_{\mathrm{s}}$. The horizontal green band shows the constraint of $\alpha_{\mathrm{s}}$ from Planck 2018 at 68\% CL. The two vertical lines denote the region of $n_{\mathrm{s}}$ from Plank 2018. From the left panel, the traditional CNI model may almost be excluded with observation. Therefore, a nonzero dissipation effect $Q \neq 0$ may be required to save the idea of the natural inflation models and consider WNI to be significant.

We now show the prediction of $\alpha_{\mathrm{s}}$ in WNI. Because $n_{\mathrm{s}}$ and $r$ in WNI have been studied in previously \cite{Reyimuaji2021JCAP,Montefalcone2023JCAP}, we already know the allowed values of model parameters that can yield $n_{\mathrm{s}}$ and $r$ consistent with observations. Following previous studies \cite{Reyimuaji2021JCAP,Montefalcone2023JCAP}, we choose the model parameters as $N=\{40,50,60\}$, $\tilde{f}=\{3.5, 5\}$, $1 \times 10^{-4} \le c_i \le 1 \times 10^{6}$ $(i=1,3)$, and $\Lambda = \{10^{16}, 10^{17}, 10^{18}\}$ GeV.

Figure \ref{FIG:WNI_linear} shows the predictions of $n_{\mathrm{s}}$, $r$ and $\alpha_{\mathrm{s}}$ in the WNI with the linear dissipative coefficient. The top half panels show the $\tilde{f}=3.5$ case, and the bottom half panels show the $\tilde{f}=5.0$ case. The left, center, and right panels show the predictions for $\Lambda = 10^{16}$, $10^{17}$, and $10^{18}$ GeV, respectively. The light gray (light green) and gray (green) regions in the $n_{\mathrm{s}}$--$r$ plots and the horizontal green band and vertical lines in the $n_{\mathrm{s}}$--$\alpha_{\mathrm{s}}$ plots show the observed constraints, which are the same as those in Fig. \ref{FIG:CNI}. The orange, blue, and red lines in all panels in Fig. \ref{FIG:WNI_linear} denote the predictions for $N=40$, $50$, and $60$, respectively. We see that WNI with a linear dissipative coefficient is successfully consistent with the observation of $n_{\mathrm{s}}$ and $r$ from the six panels in the top and third rows in Fig. \ref{FIG:WNI_linear}. Furthermore, the six panels in the second and the bottom rows in Fig. \ref{FIG:WNI_linear} show that the predicted $\alpha_{\mathrm{s}}$ in WNI with the linear dissipative coefficient can be within the allowed region. Therefore, we can conclude that WNI with a linear dissipative coefficient is a good candidate for the correct inflation model, even if we include $\alpha_{\mathrm{s}}$ in our discussion.

Figure \ref{FIG:WNI_cubic} is the same as Fig. \ref{FIG:WNI_linear} but with a cubic dissipative coefficient. As expected, WNI with the cubic dissipative coefficient is successfully consistent with the observation of $n_{\mathrm{s}}$ and $r$ from the six panels in the top and third rows in Fig. \ref{FIG:WNI_cubic}. However, from the six $n_{\mathrm{s}}$--$\alpha_{\mathrm{s}}$ panels, the allowed magnitude of the model parameters is more restricted than that of the linear dissipative coefficient case. WNI with the cubic dissipative coefficient for $\tilde{f}=3.5$ is excluded from observations because there is no combination of $n_{\mathrm{s}}$ and $\alpha_{\mathrm{s}}$ that is simultaneously consistent with these constraints from Plank data (see the three panels in the second row in Fig. \ref{FIG:WNI_cubic}). In contrast, WNI with the cubic dissipative coefficient for $\tilde{f}=5$ is consistent with observations (see the three panels in the bottom row in Fig. \ref{FIG:WNI_cubic}).

As mentioned in the introduction, the large symmetry-breaking scale $\tilde{f}$ may cause fundamental problems. In CNI, the lower limit $\tilde{f} \gtrsim 5.4$ is obtained \cite{Martin2014PDU}. Now, we are interested in determining how small $\tilde{f}$ can be in WNI. By considering $n_{\mathrm{s}}$, lower limits of $\tilde{f}$ are estimated as $\tilde{f} \gtrsim 3.8$, $4.4$, and $4.6$ for $N=40$, $50$, and $60$, respectively, at a typical cubic dissipative coefficient in WNI \cite{Reyimuaji2021JCAP}.

We consider $\alpha_{\mathrm{s}}$ and perform a numerical search of the lower limit of $\tilde{f}$. The allowed lower limits $\tilde{f} \gtrsim 4.0$, $4.0$, and $5.0$ in WNI with the cubic dissipative coefficient were determined. This finding is almost consistent with previous results \cite{Reyimuaji2021JCAP}. Our study of $\alpha_{\mathrm{s}}$ will complete the phenomenological study of WNI.

\section{Summary\label{section:summary}}
The validity of the inflation model is mainly evaluated according to the consistency of the predicted scalar spectral index $n_{\mathrm{s}}$ and the tensor scalar ratio $r$ with the CMB observations. In addition, the running scalar spectral index $\alpha_{\mathrm{s}}$ is used to further check the inflation models' consistency.

In CI scenarios, $\alpha_{\mathrm{s}}$ can be expressed in simple mathematical expressions using slow-roll parameters. In contrast, in WI scenarios, one can find exact analytical solutions for $\alpha_{\mathrm{s}}$ in principle, but one may obtain long expressions. Therefore, previous studies on WI have only shown approximate analytical solutions or numerical results for $\alpha_{\mathrm{s}}$.

This paper presents a general analytical expression of $\alpha_{\mathrm{s}}$ without approximation for the linear and cubic dissipative coefficients. In addition, we evaluate the running scalar index $\alpha_{\mathrm{s}}$ in the WNI model. The WNI (as well as CNI) model is appealing because of its well-motivated origin of the inflaton potential. We have shown that our study of $\alpha_{\mathrm{s}}$ can complete previous phenomenological studies on WNI. In particular, the lower limits of the symmetry-breaking scale in WNI became more concrete in this study.

Finally, the running of running scalar spectral index $\beta_s = d^2n_{\mathrm{s}}/d\ln k^2$ has been omitted in this study. If $\beta_s$ is included in the analysis, then the constraints change as follows:
\begin{align}
n_{\mathrm{s}} &= 0.9587 \pm 0.0056 \ (0.9625 \pm 0.0048), \nonumber \\
\alpha_{\mathrm{s}} &= 0.013 \pm 0.012 \ (0.002 \pm 0.010), \nonumber \\
\beta_s &= 0.022 \pm 0.012 \ (0.010 \pm 0.013)
\end{align}
at 68\% CL for the $\Lambda$CDM model---the Planck 2018 TT(TT,TE,EE)+lowE+lensing data \cite{PLANCK2020AA_X}. If these constraints are employed in our analysis, the numerical results in this paper will change. However, the running scalar spectral index $\alpha_{\mathrm{s}}$ must further be investigated in WI scenarios.

\section*{Acknowledgement}
The author, T. K., thanks Rudnei O. Ramos for useful discussions.

\vspace{3mm}







\end{document}